\documentclass[preprint,12pt]{elsarticle}

\usepackage{amssymb}
\usepackage[T1]{fontenc}
\usepackage[english]{babel}
\usepackage[dvipsnames]{xcolor}
\usepackage{setspace}
\usepackage{caption}
\usepackage{multirow}
\usepackage{subcaption}
\usepackage[bbgreekl]{mathbbol} 
\usepackage{amsmath}
\usepackage[ruled]{algorithm2e}
\usepackage[strings]{underscore}
\usepackage{textcomp}
\usepackage{gensymb}
\usepackage{adjustbox}

\journal{}

\begin{document}

\begin{frontmatter}

\title{Multiscale modeling of hydrogen diffusion in iron: Effect of applied stresses and dislocations.}

\author[1]{Gonzalo Álvarez}

\ead{g.alvarezm@upm.es}

\author[1]{Álvaro Ridruejo }

\ead{alvaro.ridruejo@upm.es}

\author[1,2]{Javier Segurado \corref{cor1}}

\ead{javier.segurado@upm.es}

\cortext[cor1]{Corresponding author}

\affiliation[1]{organization={Universidad Politécnica de Madrid, 
Department of Materials Science, E.T.S.I. Caminos},
addressline={C/ Profesor Aranguren 3},
postcode={28040},
city={Madrid},
country={Spain}}

\affiliation[2]{organization={IMDEA Materials Institute},
addressline={Calle Eric Kandel 2},
postcode={28906},
city={Getafe},
country={Spain}}

\begin{abstract}
Modeling hydrogen diffusion and its absorption in traps is a fundamental first step towards the understanding and prediction of hydrogen embrittlement. In this study, a multiscale approach which includes DFT simulations, OkMC, and phase-field dislocations, is developed to study the  movement of hydrogen atoms in alpha-iron crystals under the effect of applied stress and containing dislocations. At the nanoscale the interaction energies of hydrogen on different sites of the iron lattice are studied using DFT. At the microscale, this information is used to feed a lattice object kinetic Monte Carlo code (OKMC) which aims to evolve the arrangement of a large set of hydrogen atoms into the iron lattice considering point defects and the presence of dislocations. At the continuum level, an array of dislocations is introduced using a phase-field approach to accurately consider their elastic fields and core regions. The OKMC model includes both the chemical energies of H and vacancies and the elastic interactions between these point defects and the dislocations. The elastic interaction is obtained by an FFT-based approach which allows a very efficient computation of the elastic microfields created by the defects in an anisotropic medium.

The framework has been used to obtain the diffusivity tensor of hydrogen as a function of the external stress state, temperature, and the presence of dislocations.  It has been found that dislocations strongly affect the diffusivity tensor by breaking its isotropy and reducing its value by the effect of the microstresses around the dislocations.

\end{abstract}

\begin{graphicalabstract}
\begin{adjustbox}{max width=\textwidth}
\includegraphics{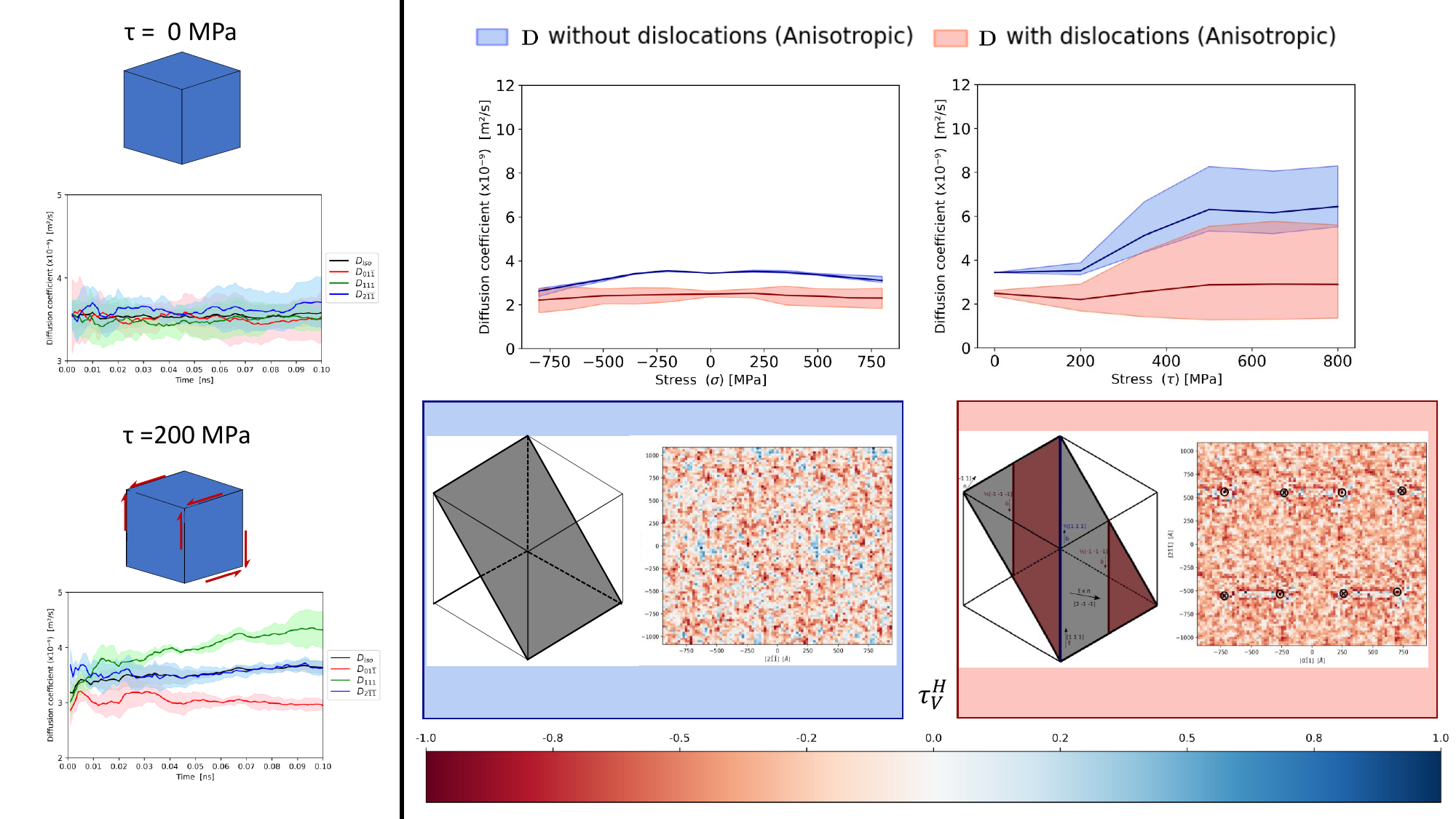}
\end{adjustbox}
\end{graphicalabstract}

\begin{highlights}
\item A multiscale model based on lattice OkMC and an FFT-based mechanical solver has been developed to study the elasto-diffusivity of hydrogen in BCC-Fe.
\item The interaction energy of hydrogen at different sites of the lattice under stress has been obtained from DFT.
\item The diffusion coefficient of hydrogen in pristine BCC-Fe have been evaluated as function of temperature and stress state.
\item The effect of dislocations on the tensorial diffusion coefficient of hydrogen in BCC-Fe has been estimated.
\end{highlights}


\end{frontmatter}



\section{Introduction}

The presence of hydrogen is related to a decrease in fracture toughness in many metallic materials, giving rise to the so-called hydrogen embrittlement. The process was first described in 1926 by Pfeil \cite{Pfeil1926} and has been profusely studied since then due to its strong technological implications. Several models have been proposed depending on the interaction and difference in relative relevance of the micro mechanisms involved, including hydrogen-induced phase transformation \cite{Motta2012, Pushilina2018}, hydrogen-enhanced decohesion \cite{Troiano1984, Dadfarnia2015}, hydrogen-enhanced localized plasticity \cite{Robertson2015, Lynch2019} and other models considering coupled mechanisms  \cite{Shishvan2023,Djukic2019}. In any case, it is clear that hydrogen embrittlement is a complex phenomenon that involves very different time and length scales, making its modeling a very challenging task.

A deep understanding of the hydrogen embrittlement process and its modeling implicates three main aspects. (1) The effect of hydrogen on the local constitutive response of the host material. (2) The distribution of hydrogen in the material under different conditions and (3) the kinetics of the evolution of the hydrogen concentration depending on the local environment.

In terms of the latter two aspects, significant efforts have been made to understand and model the process of hydrogen diffusion in different metals. At the discrete level, both density functional theory (DFT) and molecular dynamics (MD) have been used to estimate the energy landscape of H in iron, while the kinetics of H evolution has been accounted for using MD and KMC for small times \cite{Fu2005, Song2013}. On the one hand, \textit{ab initio} techniques such as DFT are able to provide accurate information of lower scales, as they provide the equilibrium electronic structure allowing to characterize the short-lived critical transition structures and configurations. Indeed, DFT has been used to quantify interaction energies and barriers for H in the lattice as well as to study the interaction with other hydrogen atoms or defects \cite{Sanchez2008, Mirzaev2014, deAndres2019, Alvarez2024}. On the other hand, the accuracy of DFT is correlated with a high computational cost that significantly limits the number of atoms in the system to some hundreds, even for very efficient computation clusters. Also at the nanoscale, a more efficient alternative is the use of MD, which is based on the computation of atomic forces using an interaction energy potential and the evolution of the system by time integration of the equations of motion. Although computationally is much cheaper than DFT the reliability of its results is based on the quality of the interatomic potentials. Standard potentials are semiempirical expressions calibrated to fit theoretical or experimental values and their accuracy can be limited since ir depends on the extrapolation to the continuum range of unknown scenarios. This limitation can be partially alleviated by informing MD potentials with DFT results to fit classical potentials \cite{Ramasubramaniam2009} or machine learning. In particular, a machine learning potential for Fe-H \cite{Zhang2022,Zhang2024} has recently been developed with very promising results. 
Nevertheless, although these techniques are much faster than rough DFT calculations, the simulations are still very expensive and can hardly reach more than $10^7-10^8$ atoms at very small times. 

Beyond atomistic approaches but preserving the discrete description, Monte Carlo methods are able to describe the system in a probabilistic manner, allowing for the reach of much higher times and length scales than atomistic models. These methods only explicitly consider the defects involved and, by defining a set of possible events with a characteristic frequency linked with their likelihood, explore the evolution of the system from a statistical viewpoint. The reliability of these methods is strongly correlated with the quality and range of description of the events involved in the simulations. Despite their potential, many of the MC approaches proposed for the evolution of H in metals employ semi-empirical potentials, \cite{Solanki11,Leyson2015,Williams2022}, without a proper calibration of energies barriers based on \textit{ab initio} methods. Moreover, even when individual events are rigorously considered, long-range elastic interactions are usually neglected. Nevertheless, this interaction is fundamental to understand the kinetics of many defects in the presence of external strain fields or microstrains induced by the other defects \cite{Clouet2008, Nazarov2016, Dudarev2018, Huang18102019}.

Carbon steels are one of the alloys susceptible to H embrittlement that has been studied in more detail because of their importance as structural materials. However, there are still many open issues regarding the H distribution in the lattice and the kinetics of its migration. One of them is how elastic strains affect H storage and migration, since these strains are generally present in an alloy as consequence of external loading, in the form of residual stresses or, at the microscale, just by the presence of other defects such as dislocations or small precipitates. As an illustration of this effect in other materials, 
Aziz \cite{Aziz2001} observed anisotropic strain-dependent diffusion in Si and SiGe for homogeneous stress fields, with an exponential dependence on the directional diffusion coefficient of three times per 1\% strain. In HSLA steels, Quan \cite{Quan1997} reported the effect of stress gradients, with a consistent increase in the diffusion coefficient with an increase in linear tensile stress gradients around 8 times per $\mathrm{TPa}\,\mathrm{m}^{-1}$.


Despite the evidence of this and other effects related to the presence of elastic strains, such as the influence of elastic dislocation distortions on H mobility, no models can be found accounting for the effect of stress on hydrogen diffusion from a microscopic perspective.


The purpose of this work is the development of a multiscale framework able to predict the diffusivity tensor of hydrogen in BCC-Fe as a function of the local stress state and temperature as well as to measure the effect of the presence of dislocations. In order to evaluate the diffusion coefficient of hydrogen, a quantum informed object kinetic Monte Carlo code \cite{Young1966}, coupled with a spectral mechanical solver \cite{Lucarini2022,Santos-Guemes2024} has been developed. This numerical framework allows us to solve the local stress distribution corresponding to the current lattice defect arrangement while performing the migration of the interstitial defects accounting for the effect of this field. The technique developed will allow to predict the effect of stress temperature and dislocation content in the diffusivity of H without introducing fitting parameters.

The paper is organized as follows. First, in section \ref{sec:okMC} the object kinetic Monte Carlo (OkMC) model used to evaluate hydrogen diffusion is described. Section \ref{sec:ContModel} present the phase field model of the different crystalline defects.  Then, section \ref{sec:Results} presents a validation of the model together with estimates of the diffusion coefficient tensor of hydrogen in iron. Finally, conclusions are drawn in section \ref{sec:Conclusions}.

\section{Multiscale simulation framework}
The multiscale simulation framework proposed to study hydrogen diffusion that accounts for the elastic interaction with defects will be presented. The core is an OkMC tool that is supported at the nanoscale by the DFT characterization of all energy barriers, considers the microscale by including the elastic interactions of hydrogen with a continuous representation of dislocations, and provides an estimate of the macroscopic diffusivity by the integration of hydrogen atoms' evolution. The different aspects of the framework will be summarized in the next sections.

\subsection{Object kinetic Monte Carlo}\label{sec:okMC}

An object lattice kinetic Monte Carlo (OkMC) code has been developed to model hydrogen diffusion through the lattice based on a first-order Markov chain. 
In this model, the system evolves following a stochastic trajectory over time. In each evolution step, some non-interacting point defects move to one of the adjacent viable positions, and a random time step following a continuous random distribution is selected.

The objects considered in the OkMC are point defects, both interstitial hydrogen and vacancies, which are allowed to move between contiguous possible sites. The probability of a certain transition, \textit{j}, to be performed by a particle, \textit{i}, at a given time, \textit{t}, is proportional to the ratio between the rate at which that step would be taken at time \textit{t} ($\nu_{ij,t}$) and the cumulative rate of all possible steps at time \textit{t} (Eq.\eqref{Eq:p_i}).

\begin{equation}
    \label{Eq:p_i}
    p_{ij,t}= \frac{\nu_{ij,t}}{ \displaystyle \sum_{i=1}^{N_i}\sum_{j=1}^{N^i_j}\nu_{ij,t} }
\end{equation}

According to transition state theory \cite{Eyring1938, Pechukas1981, Truhlar1996}, the rate at which hope \textit{i} occurs can be expressed as Eq.\eqref{Eq:nu_j}: 
\begin{equation}
    \label{Eq:nu_j}
    \nu_{ij}= \nu_{0_{ij}}\exp{\left (\frac{-\Delta E_{ij}}{k_{B}T}\right )}
\end{equation}
where the pre-exponential factor ($\nu_{0_{j}}$) is the attempt frequency of transition \textit{j} of defect \textit{i} and the exponential factor $\Delta E_{ij} / k_{B}T $ is the relationship between the energy barrier for that specific transition and the available thermal energy. 

The time between two different events is obtained randomly following a Poisson distribution with a most likely time step equal to the sum of the inverses of the frequencies of the different events. Following the expressions in Eq.\eqref{Eq:dt} and Eq.\eqref{Eq:nu_j},  where $\Re$ represents a pseudo-random number in the interval [0,1).

\begin{equation}
    \label{Eq:dt}
    \Delta t = \frac{-log(\Re)}{ \displaystyle \sum_{i=1}^{N_i}\sum_{j=1}^{N^i_j}\nu_{ij,t}}
\end{equation}

\subsection*{Energy Barrier determination}
In order to evaluate the energy barrier of a given jump, the effective energy barrier has been additively decomposed into four different terms Eq.\eqref{eq:Ebarr}. 
\begin{equation} \label{eq:Ebarr}
    E = E_0 + \Delta E_\mu + \Delta E_\sigma + \Delta E_{\nabla\sigma}
\end{equation}
Here $E_0$ is the equilibrium energy barrier arising from the transition state theory, considering a single defect that performs the transition in an otherwise pristine host lattice, under zero stress.
$\Delta E_\mu$ is the modification of the equilibrium energy barrier due to the change to the chemical potential landscape in the presence of a neighboring defect. 
$\Delta E_\sigma$ is the modification of the equilibrium energy barrier due to the local stress state. Finally, $\Delta E_{\nabla\sigma}$ is the modification of the energy barrier due to the difference in the elastic energy of the defect between the stable and transition states. Note that although having a stress origin, the terms $\Delta E_\sigma$ and $\Delta E_{\nabla\sigma}$ have a different nature. The first one considers the energy modification due to changes in the iron cell caused by elastic strains, while the second ($\Delta E_{\nabla\sigma}$) accounts for the additional change in the barrier caused the presence of elastic strain gradients.

The values for $E_0$, $\Delta E_\mu$ for two hydrogen interstitial atoms (H-IA) and one H-IA and a Fe vacancy, and $\Delta E_\sigma$ for selected stress components had previously been obtained by DFT calculations \cite{Alvarez2024}. The estimation of $\Delta E_\sigma$ for general stress states is detailed below (sec. \ref{sec:Eb_sd}), while the evaluation of the elastic contribution is described in Section \ref{sec:Elast}. 

\subsubsection*{$\Delta E_\sigma$ in complex stress states}\label{sec:Eb_sd}
The value of $\Delta E_\sigma$ has been obtained by DFT for the three uniaxial stress states in cubic directions ($[100]$) and for three pure shear cases along these planes and directions in \cite{Alvarez2024}. In order to generalize the dependencies obtained in \cite{Alvarez2024} to non-uniaxial stress states, two different approaches have been considered. The first approach, named \emph{ dominant contribution}, considers that the dependence on the stress state can be approximated to the contribution of the maximum component of the deviatoric stress on the crystal axis: $\Delta E_\sigma\left(\boldsymbol{\sigma}\right) =  \text{max}_{ij} \ \Delta E_\sigma  \left((\sigma'_{ij})\right) $. The second approach, the \emph{additive contribution}, adds the independent contribution of all components of the deviatoric stress tensor, $\Delta E_\sigma\left(\mathbb{\sigma}\right) = \sum_{i,j} \Delta E_\sigma\left(\sigma'_{ij}\right)$. 
The effect of using either approximation on the diffusivity tensor is analyzed in section 1 of the supplementary material.

\subsubsection*{Parallelization of events}
In order to accelerate the OkMC simulations the algorithm has been parallelized; to this aim, at the time of event selection, several random events are sequentially selected. Every selected event is compared with every single event selected since the last evolution. If the last selected event is farther than a threshold distance to all the previously selected events, the event is listed as being performed in the specific time increment, and the time increment is increased by the characteristic time increment of the current system. If the last selected event is close enough to any of the previously selected events, this event and its likelihood are stored for the next time increment, while the system and time are updated performing all the previously listed events. In the next time step, the new updated likelihood of the previously chosen event is evaluated and compared to its likelihood before the time step, if the new likelihood is higher, the event is listed as the first chosen event for the new time step; however, if the likelihood is smaller, the event is only listed with a probability proportional to the ratio between the likelihoods.

\subsection{Continuum model for elastic interactions between defects}\label{sec:ContModel}

In order to characterize the long range elastic interaction between an interstitial hydrogen atom (H-IA) and a secondary defect on the iron lattice (other H-IA, Vacancies (Vac), dislocations, etc.), a continuum description of the defects of the iron lattice is presented. 

Based on their dimensionality, alternative approaches have been taken to describe the different defects:

\subsubsection{Dislocations}\label{sec:Disloc}

The elastic fields generated by dislocations are described using a phase-field dislocation (PFD) framework \cite{Beyerlein2016}. In this approach, a dislocation loop with burgers vector $\mathbf{b}$ and normal \textbf{n} is represented by an smooth field (order parameter $\phi$) defined on its own plane. The order parameter $\phi$ represents the relative displacement in the direction of the burgers vector between the upper and lower parts of the loop, normalized by the burgers vector module. In PFD, the order parameter defines an eigenstrain in a very thin region $h$  (ideally, an atomic plane) which is the relative displacement divided by $h$, Eq.\eqref{eq:disloc_eig}
\begin{equation}\label{eq:disloc_eig}
   \boldsymbol{\varepsilon}^{Eig}(\phi^\alpha,\mathbf{x})=\frac{(\mathbf{b}^\alpha\otimes\mathbf{n}^\alpha)+(\mathbf{n}^\alpha\otimes\mathbf{b}^\alpha)}{2h}\phi^\alpha(\mathbf{x})=\boldsymbol{\Lambda}^\alpha\phi^\alpha(\mathbf{x})
\end{equation}

The Helmholtz free energy of a body with a dislocation is given by a functional $\Pi[\phi]$ (Eq. \eqref{eq:func_PFD}) which includes the bulk energy density, composed by the sum of the elastic, $\Psi_e$, and the lattice, $\Psi_l$, contributions and the energy associated to the phase field gradient,  $\Psi_g$,  that represents the core of the dislocation,  
\begin{equation}   \label{eq:func_PFD}
      \Pi\left[\phi^\alpha\right]=\int_\Omega
      \Psi_e\left[\phi^\alpha(\mathbf{x})\right]+
      \Psi_l\left[\phi^\alpha(\mathbf{x})\right]+
      \Psi_g\left[\nabla\phi^\alpha(\mathbf{x})\right]
      \mathrm{d}\Omega.
\end{equation}
The elastic energy in a homogeneous linear elastic medium with stiffness $\mathbb{C}(\mathbf{x})=\mathbb{C}$ is given by a quadratic function of the elastic strain. Assuming stress equilibrium by the conservation of linear momentum, this energy can be expressed as function of the phase field through its eigenstrain (Eq. \eqref{eq:disloc_eig}) convoluted with the Green's function second derivative $\mathbb{\Gamma}$, resulting in
\begin{equation}  
    \Psi_e(\mathbf{x},\phi^\alpha)=\frac{1}{2} \mathbb{C}:\boldsymbol{\varepsilon}^{e}:\boldsymbol{\varepsilon}^{e}
\end{equation}
where the elastic strain is 
\begin{equation}\label{eq:el_strain}
\boldsymbol{\varepsilon}^{e}=-\left(\mathbb{\Gamma}*(\mathbb{C}:\boldsymbol{\varepsilon}^{Eig})-\boldsymbol{\varepsilon}^{Eig} \right)
\end{equation}
A description of the solution of the elastic problem in the presence of eigenstrains in an FFT framework is provided in section 2 of the supplementary material.

The lattice energy density associated with a slip system, $\Psi_l$ is well approximated in BCC iron by a sinusoidal potential of height $U$ \citep{Hu2021},  $U = \gamma/h $ with $\gamma = 0.944 \mathrm{J/m}^2 $ for the slip system $\frac{1}{2} \left< 1 1 1 \right> \{ 1 1 0 \}$,
\begin{equation}   
      \Psi_l(\mathbf{x},\phi^\alpha )=U \sin^2(\pi \phi^\alpha(\mathbf{x}) ).
\end{equation}
Finally, the gradient contribution smears the dislocation core penalizing the gradient of the PF along the dislocation plane and is correlated to the dislocation core energy. This energy term is given by 
\begin{equation}   
       \Psi_g(\mathbf{x},\phi^\alpha)= G|\mathbf{b}|^2 \sum_\alpha  [(\mathbf{n}^{(\alpha)}\times \nabla) \phi^\alpha(\mathbf{x})]^2.
 \end{equation}
with $G$ the shear modulus in plane $\mathbf{n}$ along burgers vector direction. 

To represent a dislocation loop in a static simulation, an initial non-smooth phase field function (with value 1 in the loop interior) is introduced to define the loop shape. Then, a relaxation of the total energy functional  (Eq. \eqref{eq:func_PFD}) is performed with the boundary conditions corresponding to the specific simulation. The resulting phase field and elastic fields (Eq. \eqref{eq:el_strain}) are used as a fixed background state for the inclusion of point defects. 
The numerical implementation of the PFD approach in a FFT framework is detailed in section 3 of the supplementary material.

\subsubsection{Point defects}\label{sec:PointDef}
Point defects, such as hydrogen interstitial atoms (H-IA) and vacancies (Vac), have been described as a parallelepipedal inclusion of the host matrix containing the defect on a homogeneous medium of pristine BCC-Fe, and can be characterized by their dipole approximation, the first moment of their characteristic point force distribution \cite{Clouet2018}.


The displacement field associated with a point defect in a solid can be modeled as the resultant displacement field for a distribution of point forces under equilibrium around the defect \cite{Bacon1980}. Defining a dipole tensor $\mathbf{P}$ as the first moment of the equivalent force distribution, Eq. \eqref{eq:dipole_app}, this displacement at a  position $\mathbf{r}$ from the point defect corresponds to
\begin{equation}\label{eq:dipole_app}
    u_i(\mathbf{r}) = - G_{ij,k}(\mathbf{r}) P_{jk}
\end{equation}
where $G_{ij}$ is the Green's function of the elastic medium.

The dipole approximation can also be considered by either the relaxation volume tensor($\boldsymbol{\Omega}$) or the $\lambda$-tensor ($\boldsymbol{\lambda}$), which are conjugated to the dipole tensor $\mathbf{P}$, by the matrix fourth order stiffness tensor $\mathbb{C}$ and the defect relaxation volume $\Delta V$,
\begin{equation}
    \mathbf{P} = \mathbb{C} : \boldsymbol{\Omega} = \Delta V\mathbb{C}: \boldsymbol{\lambda} 
\end{equation}

The dipole tensor can be obtained from \textit{ab initio} calculations using the \textit{stress} or the \textit{strain methods} \cite{Clouet2008, Nazarov2016, Dudarev2018}, which can be attributed to a completely unresponsive and completely responsive matrix. In this work, due to the characteristic speed of the relevant transitions ($\approx 10^{3}$m/s), being about 20\% the speed of sound in steels, the strain method was considered. The dipole tensor of an interstitial hydrogen atom in the stable, $\mathrm{Fe}_{54}H^{T}$, and first $\mathrm{Fe}_{54}H^{X}$ and second $\mathrm{Fe}_{54}H^{O}$ degree saddle points, and an iron vacancy $\mathrm{Fe}_{53}$ in the BCC-Fe lattice have been calculated using CASTEP (v.22). On-the-fly generated pseudopotentials under the PBE formulation with partial core correction for iron were used in the calculations. The cutoff energy of the plane wave basis set used was 375 eV with finite basis set correction, with a k-mesh following a regular MP arrangement of 4x4x4 without offset for a calculation containing a supercell of 3x3x3 BCC-Fe unit cells (V=600.955 \AA$^3$). All the calculations performed in this work were spin polarized. The reference configuration for the calculations was the pristine BCC cell without the inclusion of the point defect. The elastic stiffness used for all dipole tensor calculations was evaluated for the reference configuration.

Table \ref{tab:omega_vals} shows the relaxation volume tensor for the single point defect configurations studied when the defect is aligned to the $\mathbf{x}$ direction, which means that the dissimilar direction of the symmetry matrix of the cell containing the defect is aligned to the $\mathbf{x}$ direction, while the $\mathbf{y}$ direction and the $\mathbf{z}$ direction directions are equivalent \cite{Alvarez2024}.
\begin{table}[ht]
    \centering
    \begin{tabular}{|c|c|c|c|c|}
        \hline
         Point defect & $\Omega_{xx}$ & $\Omega_{yy}$ = $\Omega_{zz}$ & $\Omega_{xy} = \Omega_{xz}$ & $\Omega_{yz}$ \\
         \hline
         \hline
         $\mathrm{Fe}_{54}H^{T}$ & 13.336 & -2.25625 & 0 & 0 \\
         \hline
         $\mathrm{Fe}_{54}H^{X}$ & 6.50616 & -2.37856 & 0 & 0.648097 \\
         \hline
         $\mathrm{Fe}_{54}H^{O}$ & 5.92855 & -0.627972 & 0 & 0 \\
         \hline
         $\mathrm{Fe}_{53}$ & -0.930901 & -0.930901 & 0 & 0 \\
         \hline
    \end{tabular}
    \caption{Relaxation volume tensor ($\boldsymbol{\Omega}$) components for the studied point defects in BCC-Fe aligned to the x direction in \AA$^3$}
    \label{tab:omega_vals}
\end{table}

The elastic strain created by a point defect can be obtained taking the gradient of Eq. \eqref{eq:dipole_app}. In the case of an isotropic medium, analytical expressions can be derived. However, in the case of a non-isotropic three dimensional medium (as a cubic crystal) closed expression are not possible requiring to solve numerically an eigenvalue problem \cite{SALES1998247}, losing therefore the simplicity of the approach. For this reason, and in order to use the same method to obtain elastic field distributions of both point defects and dislocations, in this work a numerical approach based on Eshelby method is used. To this aim, the volume relaxation tensor, $\boldsymbol{\Omega}$ at a position, $\mathbf{X}^d$, is distributed as an eigenstrain in a very small but finite volume. The spreading is done on a spherical region with a characteristic length $\ell= 0.5$\AA  $(\approx1/6 \,\mathrm{a}_{\mathrm{Fe}}$) following a three-dimensional exponential decay kernel, $\Xi(\mathbf{x})$, to reduce numerical noise. The resulting smooth eigenstrain fields are used to evaluate the stress field of each specific oriented point defect in the iron matrix.
\begin{equation} \label{eq:dip_eig}
    \boldsymbol{\varepsilon}^{Eig}(\mathbf{x} -\mathbf{X}^d)
    =    
     \frac{\boldsymbol{\Omega}}{V_\ell } \Xi \ast
     \delta(\mathbf{x} -  \mathbf{X}^d)
    =
    \boldsymbol{\lambda} \phi^d(\mathbf{x} -\mathbf{X}^d)
\end{equation}
In Eq. \eqref{eq:dip_eig}, $\delta$ represents the Dirac delta function and $\mathrm{V}_\ell$ is the volume of the sphere in which the eigenstrain is concentrated. The strain field is finally represented by the product of the defect tensor $\boldsymbol{\lambda}$ and $\phi^d$, which is an scalar smooth decay function, result of the convolution. 
The numerical implementation of the convolution and obtention of the function  $\phi^d$ is detailed in section 4 of the supplementary material.

\subsubsection{Elastic interaction energy}\label{sec:Elast}

The elastic energy in a domain $\Omega$ containing some elastic field ($\boldsymbol{\sigma}(\mathbf{x}),\boldsymbol{ \varepsilon}^e(\mathbf{x})$) is given by Eq.\eqref{eq:Elast_E}
\begin{equation}
\label{eq:Elast_E}
 \Delta E = \int_{\Omega} \frac{1}{2} \boldsymbol{\sigma}(\mathbf{x}) : \boldsymbol{\varepsilon}^e(\mathbf{x}) = \int_{\Omega} \frac{1}{2} \mathbb{C}(\mathbf{x}) : \boldsymbol{\varepsilon}^e(\mathbf{x}): \boldsymbol{\varepsilon}^e(\mathbf{x}).
\end{equation}
The elastic interaction energy of a system with two elastic fields, $\Delta E^{1,2}$, can be calculated as the difference between the elastic energy of a system with a superposition of both fields and elastic energies of the systems containing the individual fields (Eq.\eqref{eq:IntE}) \cite{Santos-Guemes2024, Alvarez2025Zr}.

\begin{equation}\label{eq:IntE}
    \begin{aligned}
        &\Delta E^{1,2} = E^{1 \cup 2} - E^{1} - E^{2} =
        \int_{\Omega} \boldsymbol{\sigma}^{1} : \boldsymbol{\varepsilon}^{2} =
        \int_{\Omega} \boldsymbol{\sigma}^{2} : \boldsymbol{\varepsilon}^{1}
    \end{aligned}
\end{equation}

In order to evaluate the elastic interaction energy between a defect and the surrounding system, it is sufficient to evaluate the elastic interaction energy between the elastic field attributed to the defect and the elastic field present in the system (excluding the defect field).

To this aim, the stress fields of the point defects considered are computed at the beginning of the simulation on a pristine lattice at the origin and stored in their frequency space representation. These fields are translated to the positions of their corresponding defects, $\mathbf{X}^{d}$, using the shift theorem, as proposed in \cite{Santos-Guemes2024, Alvarez2025Zr}. 

Calculating the difference in elastic interaction energy between the stable defect \textit{i} in its stable position, $\mathbf{X}_0^i$ and its saddle point defect, in its saddle point position for the transition \textit{j}, $\mathbf{X}_j^i$, with a background strain field, $\varepsilon^{Sys}(\mathbf{x})$ can be done using Eq.\eqref{eq:Int_dif}
\begin{equation}\label{eq:Int_dif}
    \begin{aligned}
        &\Delta^{j,0}E =   E^{j,Sys} -  E^{0,Sys} = \\
        &\int_{\Omega} \boldsymbol{\sigma}(\mathbf{X}_j^i;\mathbf{x}) : \boldsymbol{\varepsilon}^{Sys} (\mathbf{x}) 
        -        
        \int_{\Omega} \boldsymbol{\sigma}(\mathbf{X}_0^i;\mathbf{x}): \boldsymbol{\varepsilon}^{Sys} (\mathbf{x})
        =        \\
        &\int_{\Omega} \left( \boldsymbol{\sigma}(\mathbf{X}_j^i;\mathbf{x}) - \boldsymbol{\sigma}(\mathbf{X}_0^i;\mathbf{x}) \right) : \boldsymbol{\varepsilon}^{Sys} (\mathbf{x}) 
    \end{aligned}
\end{equation}
The evaluation of Eq. \ref{eq:Int_dif} is done by full numerical integration of the fields obtained by the FFT solver, as shown in section 2 of the supplementary material.
\section{Results}\label{sec:Results}

The use of the full elastic strain fields of point defects and dislocations in the calculation of elastic interactions is compared with analytical expressions based on dipoles, providing almost identical results. In this work, the full fields have been evaluated using a periodic regular grid of $128^3$ voxels. The comparison is presented in section 5 of the supplementary material. 

\subsection{Hydrogen diffusivity}\label{sec:Dif}
\begin{figure}[ht]
    \centering
    \includegraphics[width=0.5\linewidth]{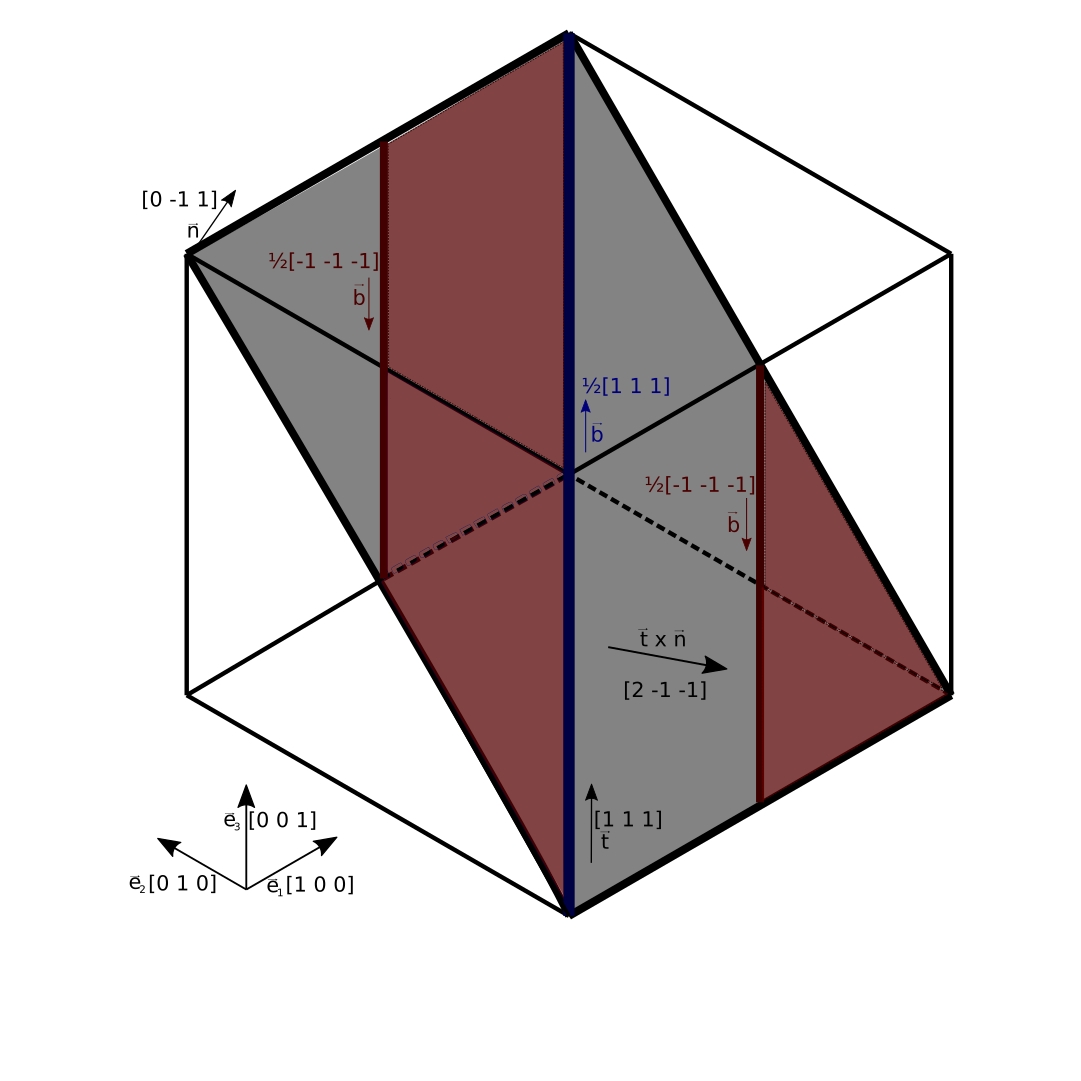}
    \caption{Schematic representation of the simulation cell including a dislocation array. The sheared area between opposite dislocations ($\phi \approx 1$) is colored in red.}
    \label{fig:scheme}
\end{figure}

Directional hydrogen diffusion in iron under different external stress states and temperatures has been evaluated using Einstein's equation for diffusion \cite{Einstein1905}. In this expression, Eq.\eqref{Eq: D_1D}, the diffusivity in a direction $\mathbf{u}$, $D_\mathbf{u}$, is obtained as the average of the square of the displacement of each solute hydrogen ($\mathbf{X}(t)-\mathbf{X}_0$) in that direction for long times. 
\begin{equation}\label{Eq: D_1D}
    D_{\mathbf{u}}=\lim_{t \to \infty}\left< \frac{\big(\left(\mathbf{X}(t)-\mathbf{X}_0\right) \cdot \mathbf{u}\big)^2}{2 \ t}\right>
\end{equation}
The diffusion coefficient in  Eq.\eqref{Eq: D_1D} has been evaluated from the displacements of hydrogen atoms obtained in OkMC simulations for at least 0.1 ns (until stabilization).
Directions have been defined on a Cartesian basis following the crystalline directions: $e_1=[100], e_2=[010], e_3=[001]$, as depicted in Fig.\ref{fig:scheme}

\subsubsection{H diffusivity as function of temperature and stress}\label{sec:Dif_homo}

The diffusion coefficient has been evaluated for temperatures between 240 and 360 K and applied external simple stress states of up to 800 MPa, which corresponds to typical operation or residual stresses in steels. All simulations shown are performed in a cubic cell with side, L = 463 $a_{\alpha-Fe}$ (L$\approx 0.136\mu$m). A hydrogen atomic concentration of $3 \times 10^{-6}$, and an iron vacancy concentration of $1 \times 10^{-7}$ are used, standard values of thermodynamical defect equilibrium. The initial positions for both species were randomly set. An attempt frequency $\nu_0^H$ of 1 THz was used for the H transitions, while the attempt frequency for the movement of the vacancies, $\nu_0^{Vac}$, considered was one order of magnitude lower, values taken from \cite{Bombac2017}.  
Note that, even in the case where external stress is not considered, internal microstresses originated by the point defects are considered and modify the effective barrier for defect jumps.

\begin{figure}[ht]
    \centering
    \includegraphics[width=0.48\linewidth]{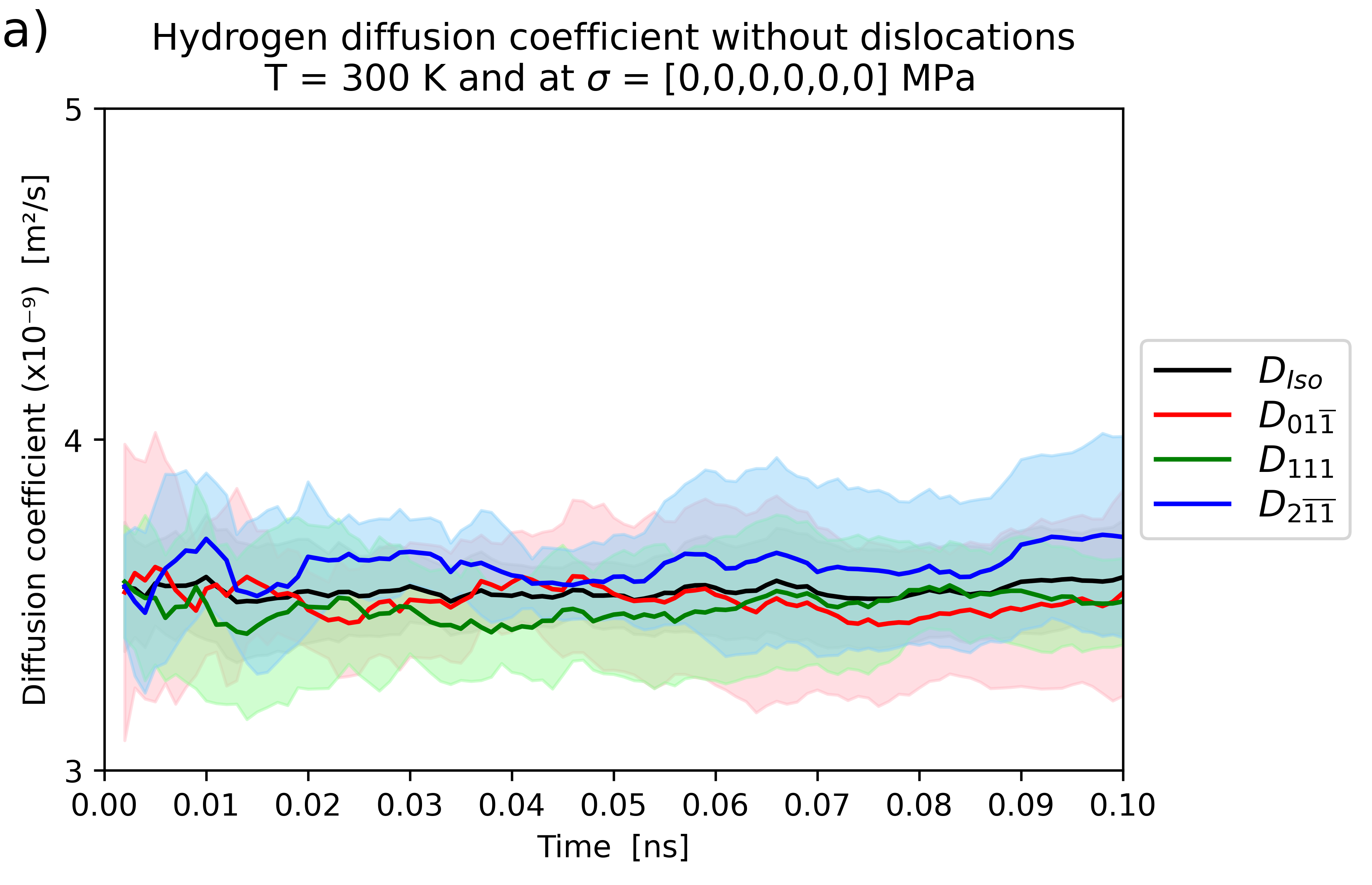}
    \includegraphics[width=0.48\linewidth]{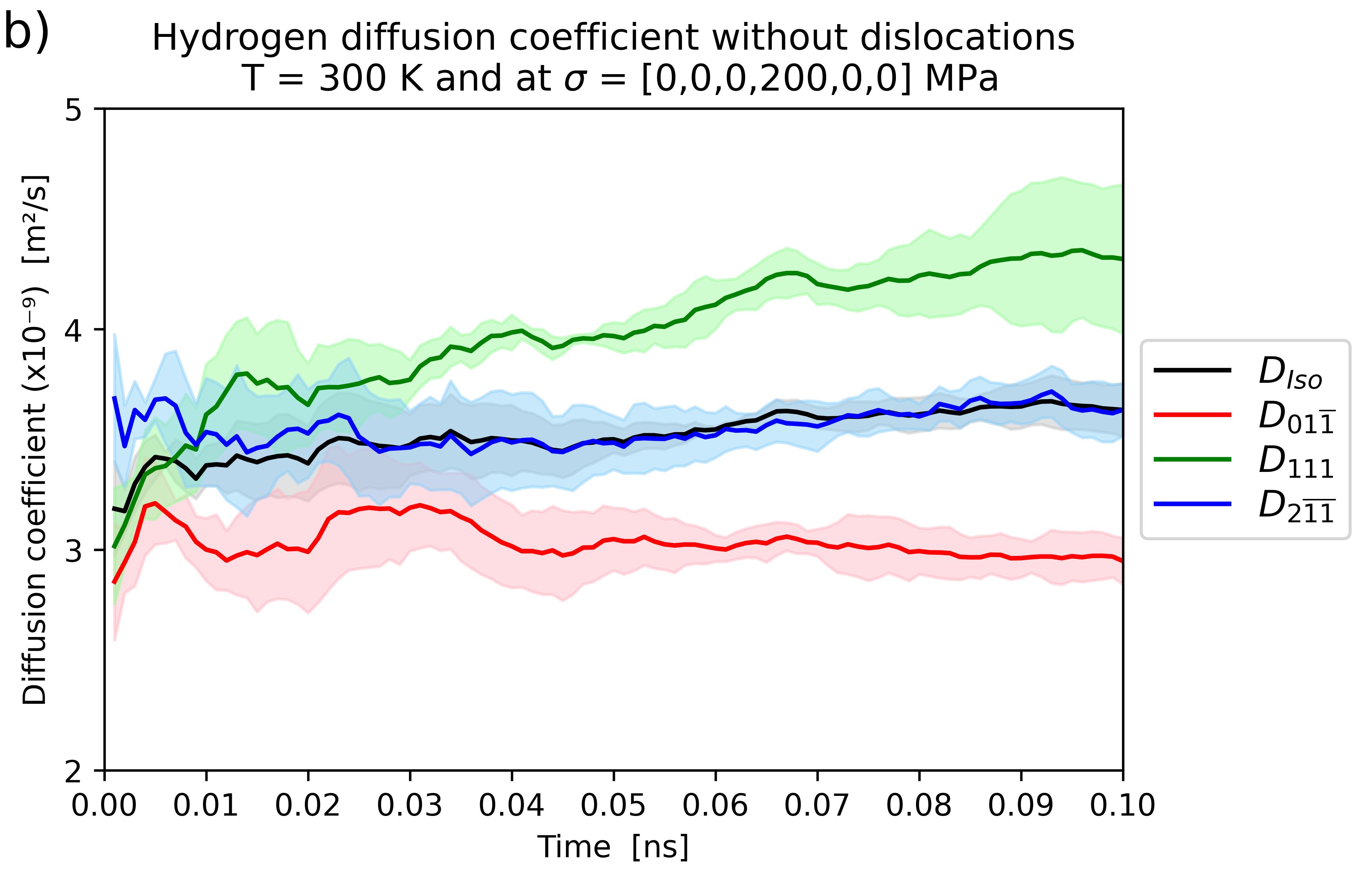}
    \caption{Evolution over time of the mean and directional diffusion coefficient of hydrogen in iron at T =300 K. a) Under no external stress. b) Under shear external stress state \\ $\boldsymbol{\sigma} = 200 (e_2 \otimes e_3)$ MPa.}
    \label{fig:D_t_evo}
\end{figure}

To illustrate the results, Figure \ref{fig:D_t_evo} shows the evolution of the mean and directional diffusion coefficients of hydrogen for two sets of simulations at T=300K, one set without external stresses and a second one under a shear stress of 200MPa along the cubic plane $(100)$. The Fig. \ref{fig:D_t_evo} comprises the results of 9 simulations for each case, and represents with lines the average diffusion obtained from the different simulations and with shadow colored areas the dispersion.
Comparing the two curves in Fig.\ref{fig:D_t_evo} it can be observed that the presence of an external applied stress not only modifies the average diffusion coefficient but also introduces an anisotropic behavior.

The results for all the temperatures considered in the absence of external stress are summarized in Table \ref{tab: D param no S}, and compared with experimental and theoretical values reported in the literature. In the table, three values are obtained with the method proposed, one without considering the effect of stress in the energy barrier (Num-no stress), and other two considering the two approaches used to introduce the effect of non-unixial stress (\emph{Num-Sup} and \emph{Num-Dom} for the superposition and dominant approaches respectively). Assuming that the diffusivity tensor $\mathbf{D}$ follows an Arrhenius dependency with temperature as
\begin{equation}
    \mathbf{D}=\mathbf{D}^0 \exp\left (\frac{-\Delta E}{k_B T}\right)
    \label{eq:arrhenius}
\end{equation}
where  $\mathbf{D}^0$ is the prefactor and $\Delta E$ the effective energy barrier, 
the table include also both the values of isotropic diffusivity prefactor $D^0=\frac{1}{3} \mathrm{tr}\mathbf{D}^0$ and $\Delta E$.

\begin{table}[ht]
    \centering
    \begin{tabular}{|c|c|c|c|c|}
       \hline
       Study & Method & D$_{(300\mathrm{K})}$ &  $\mathrm{D}^0$ & $\Delta \mathrm{E}^B$ \\
        &  &  $ \times 10^{-9} \mathrm{m}^2/\mathrm{s}$ &  $ \times 10^{-9} \mathrm{m}^2/\mathrm{s}$ &  meV\\
       \hline
       \hline
       Beck \textit{et al.} \cite{Beck1966}  & EC  & 6.46  & 60  & 57.6\\
       \hline
       Oriani \textit{et al.} \cite{Oriani1970}  & G &  3.23 &  78 & 82\\
       \hline
       Choi \textit{et al.} \cite{Choi1970}  & EC & 1.21  &  220 & 135\\
       \hline
       Heumann \& Domke \cite{Heumann1972}  & G &  4.45 & 47.4  & 61\\
       \hline
       Asano \textit{et al.} \cite{Asano1973}  & EC & 7.4  & 150  & 78\\
       \hline
       Quick \& Johnson \cite{Quick1978}  & G &  0.95 &  16.1 & 73\\
       \hline
       Hagi \textit{et al.}  \cite{Hagi1979}  & EC &  7.50 & 110  & 69\\
       \hline
       Nagano \textit{et al.} \cite{Nagano1981}  & EC & 9.10  & 220 & 135\\
       \hline
       Nagano \textit{et al.} \cite{Nagano1982}  & G \& EC &  8.7 &  42.0 & 40\\
       \hline   
       Hayashi \textit{et al.} \cite{Hayashi1989}  & G + EC  & 8.65  &  33.5 & 35\\
       \hline
       Hagi \textit{et al.} \cite{Hagi1994}  & G &  9.56 &  58 & 46.6\\
       \hline
       Jiang \textit{et al.} \cite{Jiang2004} & Num, $\Gamma_0$ & 4.99 & 150 & 88.0\\
       \hline
       Jiang \textit{et al.} \cite{Jiang2004} & Num, $\Gamma_0(T)$ & 8.66 & 44.0 & 42.0\\
       \hline
       This study & Num, Sup  & 3.53  & 34.92  & 59.3\\
       \hline
       This study & Num, Dom  & 3.24  &  29.68 & 57.2 \\
       \hline
       
    \end{tabular}
    \caption{Comparison of the diffusivity parameter with the literature. The \emph{method} column includes Electrochemical permeation experiments (EC), Gas permeation experiments (G) and theoretical values obtained from simulations (Num). }
    \label{tab: D param no S}
\end{table}

As shown in Table \ref{tab: D param no S}, and consistent with the compilation of tests performed by Hayashi and Shu \cite{Hayashi2000} there exists a wide range of experimental measurements for hydrogen diffusivity on BCC-Fe, with values for $\Delta \mathrm{E}$ and $\mathrm{D}^0$ ranging from 35 to 142 meV and from 16.1 $\times 10^{-9}$ to 223 $ \times 10^{-9}$ $m^2/s$, respectively. The values reported by our predictions fall into this experimental range. In the case of the diffusivity at room temperature, it falls in the lower range of the experimental values. 

Note that the results of the OkMC simulations are linearly dependent with the value of the event frequency used, $\nu_0$, here taken $\nu_0^H=1 $THz and $\nu_0^{Vac}=0.1$THz following typical values used in the literature. There is no clear consensus about this value. Theoretical estimations provide values at high temperature of $\nu^0 = k_BT/h \approx6.21$ THz (with $h$ Plank's constant) and of $\nu^0 =\approx 6.93 \,\mathrm{THz}$  for very low temperatures \cite{Vineyard1957, Katz1971}, but the temperature range considered lies between these states and the dependency of frequency with temperature is not monotonic. Therefore, the absolute values of diffusivity obtained have a small uncertainty, result of the uncertainty in the attempt frequency.

\begin{figure}[ht!] 
    \begin{adjustbox}{max totalsize={\textwidth}{0.26\textheight}}
    \begin{subfigure}[b]{0.50\textwidth}
         \centering
         \includegraphics[width=\textwidth]{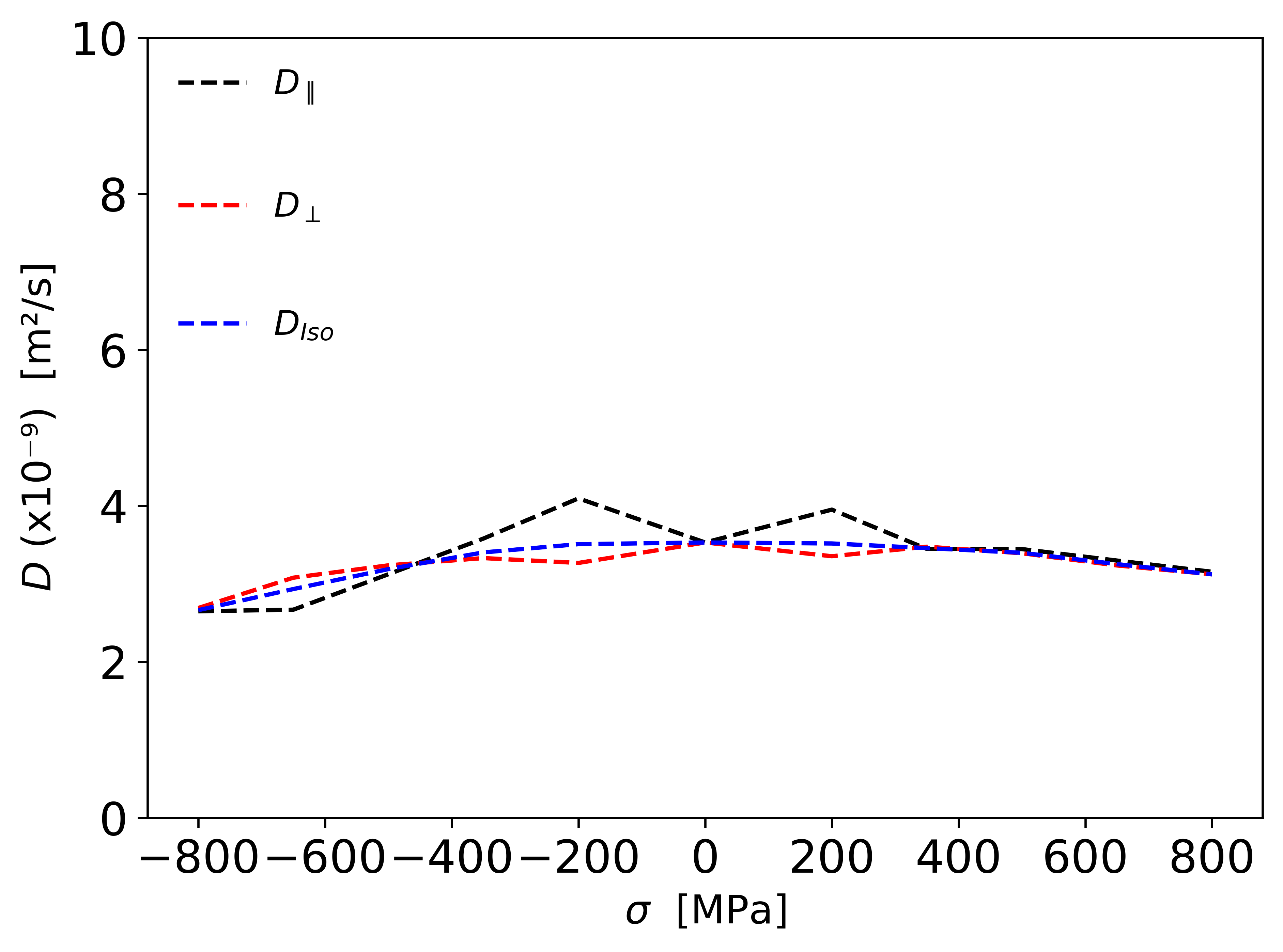}
         \subcaption[]{}
    \end{subfigure}
    \begin{subfigure}[b]{0.50\textwidth}
         \centering
         \includegraphics[width=\textwidth]{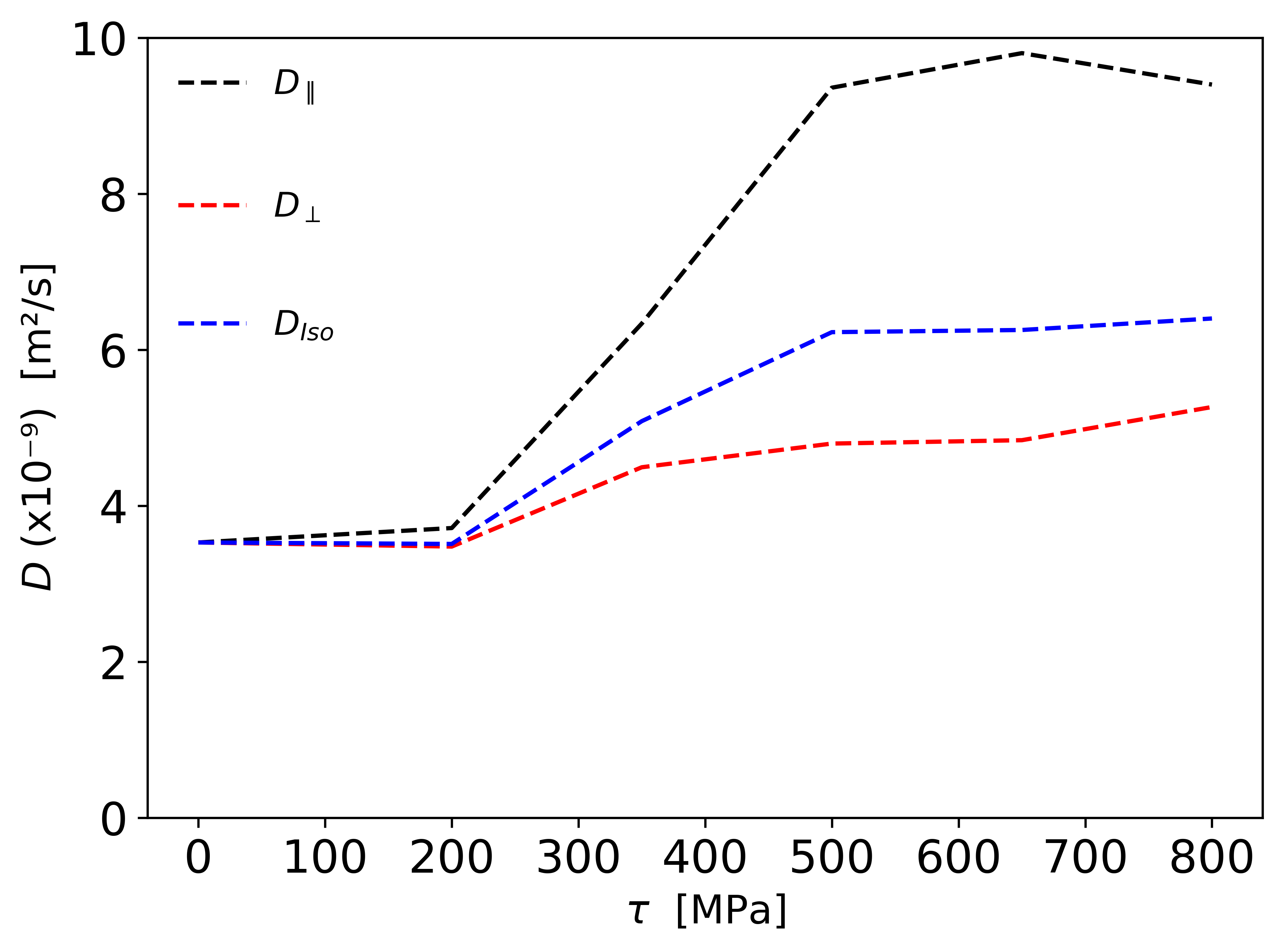}
         \subcaption[]{}
    \end{subfigure}
    \end{adjustbox}
    \begin{adjustbox}{max totalsize={\textwidth}{0.26\textheight}}
    \begin{subfigure}[b]{0.50\textwidth}
         \centering
         \includegraphics[width=\textwidth]{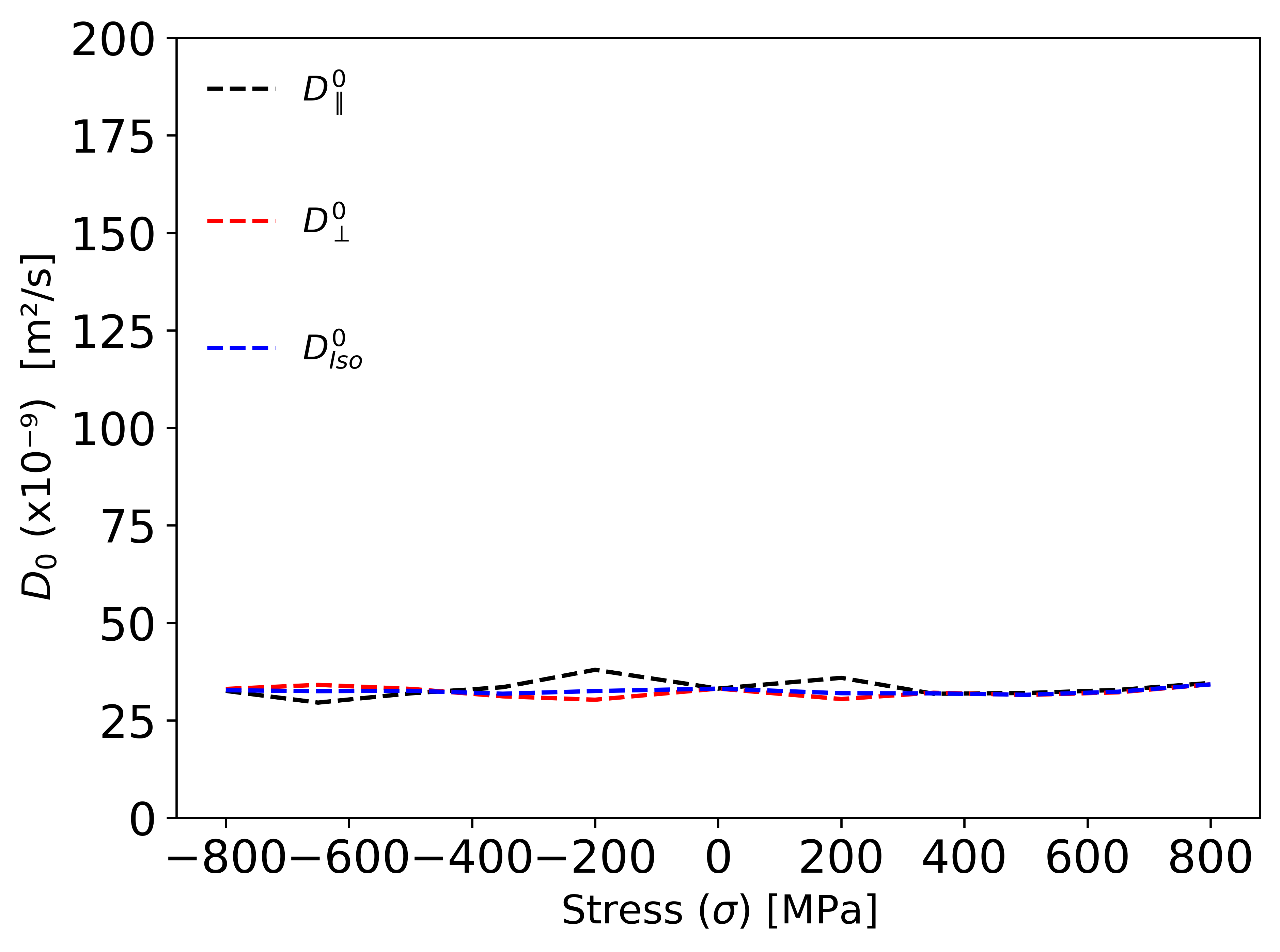}
         \subcaption[]{}
    \end{subfigure}
    \begin{subfigure}[b]{0.50\textwidth}
         \centering
         \includegraphics[width=\textwidth]{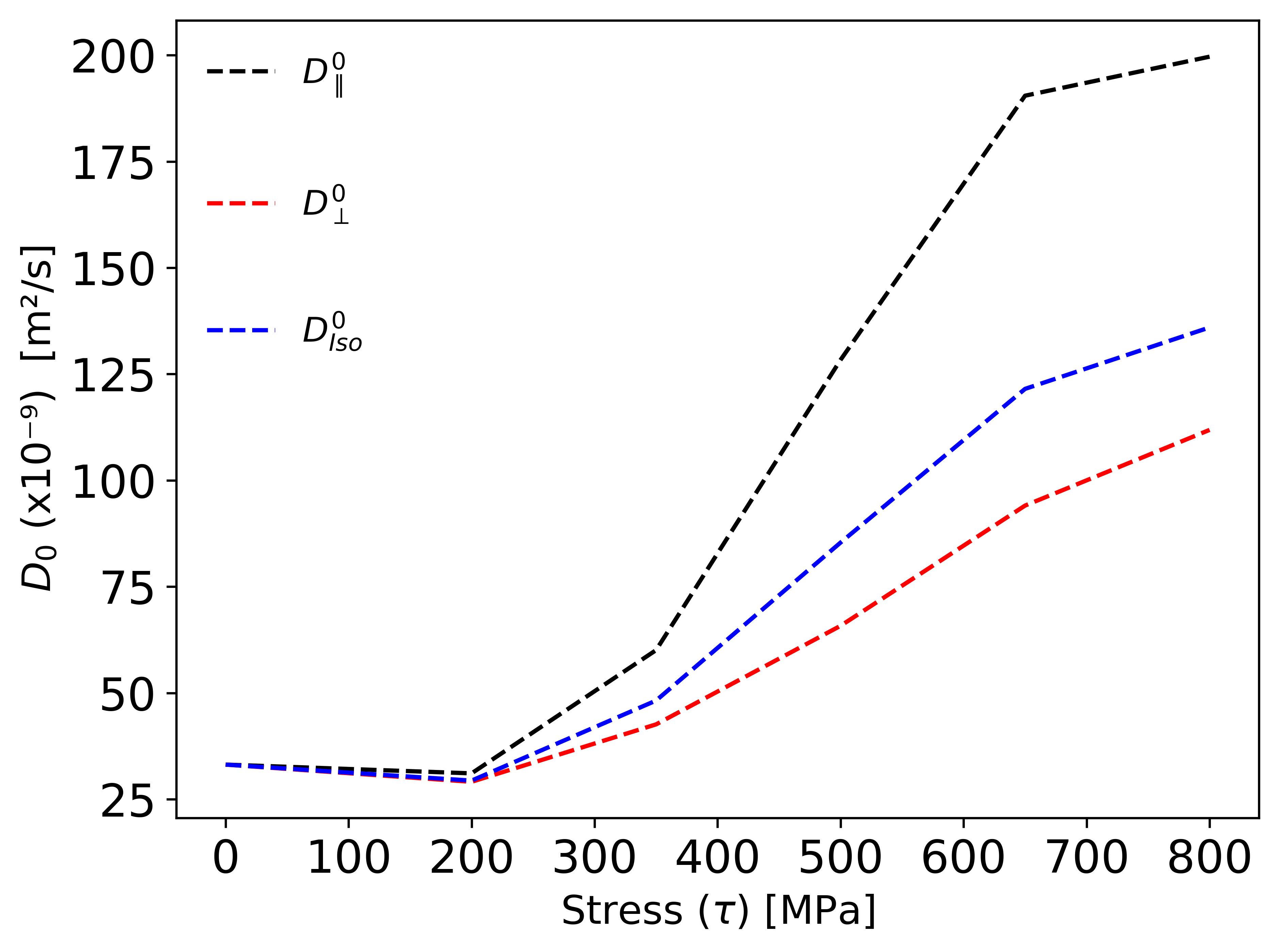}
         \subcaption[]{}
    \end{subfigure}
    \end{adjustbox}
    \begin{adjustbox}{max totalsize={\textwidth}{0.26\textheight}}
    \begin{subfigure}[b]{0.50\textwidth}
         \centering
         \includegraphics[width=\textwidth]{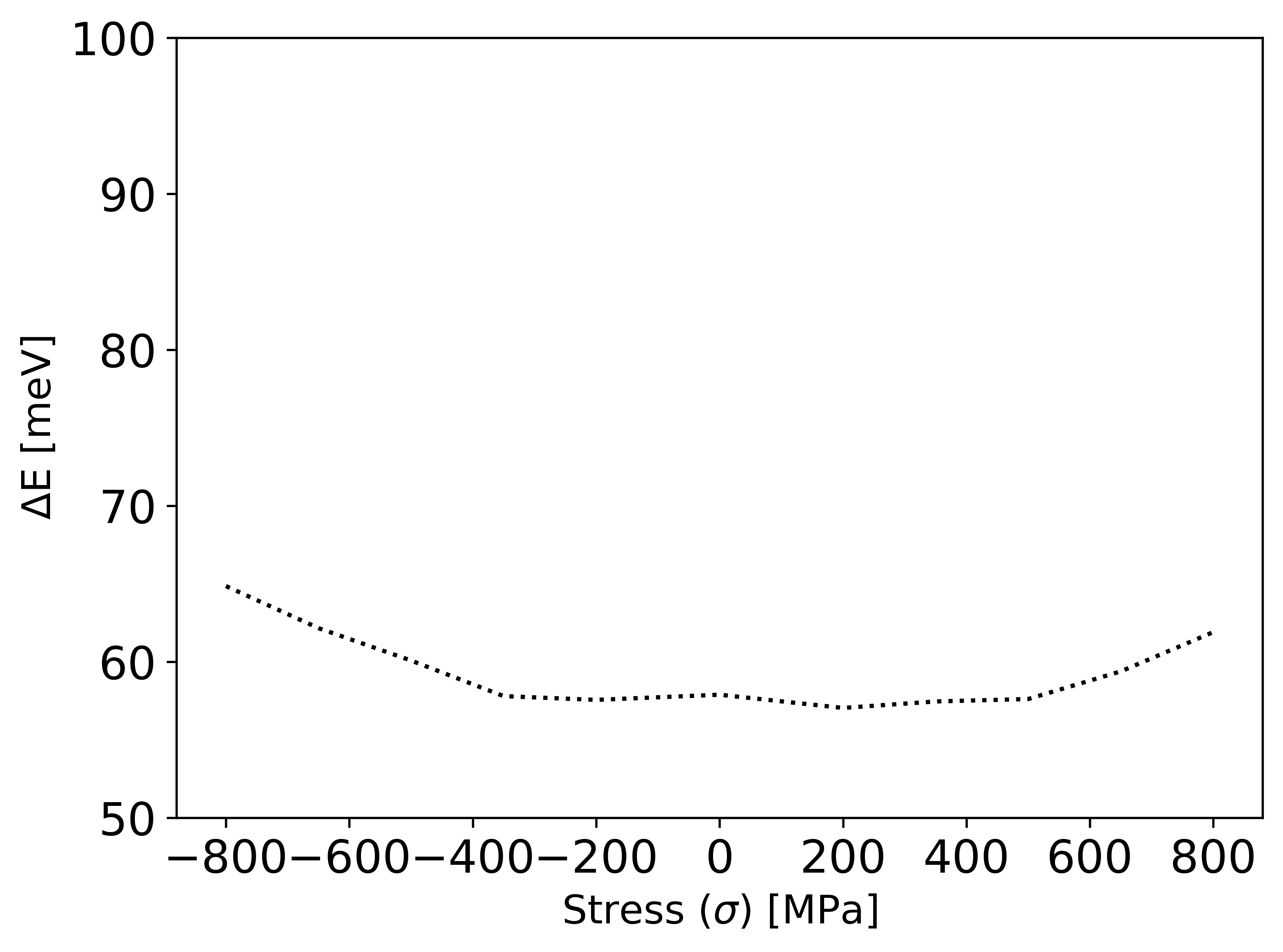}
         \subcaption[]{}
    \end{subfigure}
    \begin{subfigure}[b]{0.50\textwidth}
         \centering
         \includegraphics[width=\textwidth]{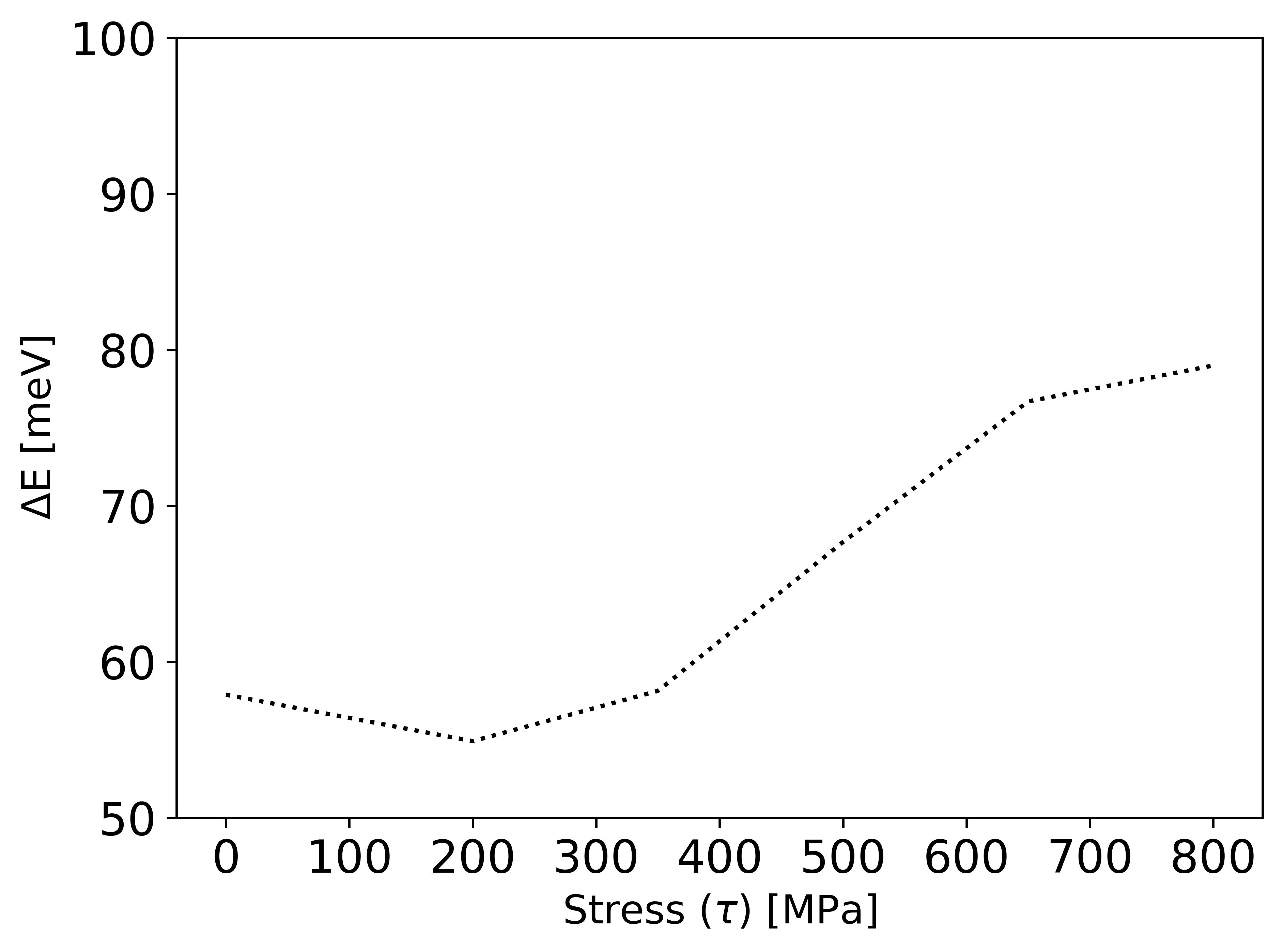}
         \subcaption[]{}
    \end{subfigure}
    \end{adjustbox}
    \caption{Stress induced anisotropy in the diffusion coefficient parameters: 
    Diffusion coefficient at 300 K (D) (a,b), pre-exponential coefficient (D$^0$) (c,d) and effective energy barrier ($\Delta E$) (e,f). Under uniaxial stress (a,c,e) and shear stress (b,d,f).
    Note that in the uniaxial cases (a,c,e), the parallel ($\parallel$) and perpendicular ($\perp$) directions refer the direction parallel to the external load, and any direction perpendicular to it respectively. However, in the shear cases (b,d,f), the parallel and perpendicular attributes are related to the normal direction to the shear plane instead.
    }
    \label{fig:D params nd}
\end{figure}

The introduction of an external stress result in two main effects, the development of anisotropy in the diffusion and the modification of the mean diffusion coefficient. Considering an Arrhenius expression also in the presence of external stress, the dependence of the diffusion tensor on both elastic stress and temperature corresponds to
\begin{equation}
    \mathbf{D}(\boldsymbol{\sigma},T)=\mathbf{D}^0(\boldsymbol{\sigma}) \exp\left (\frac{-\Delta E(\boldsymbol{\sigma})}{k_B T}\right).
    \label{eq:arrhenius2}
\end{equation}
To illustrate this effect, the diffusion coefficient at 300K, the prefactor and the effective barrier of Arrhenius expression (Eq. \eqref{eq:arrhenius2}) are represented in Fig.\ref{fig:D params nd} as function of the applied stress for uniaxial and a pure shear cases. The anisotropy induced in the diffusion tensor when external stresses are applied can be observed
in Fig.\ref{fig:D params nd} by comparing the diffusivity differences in the loading direction $D_\||$ or perpendicular to it $D_\perp$. For uniaxial stresses (Figs.\ref{fig:D params nd}(a) and (c) ) the effect is minimal and the difference between the different directions is comparable to the statistical scatter. On the contrary, the effect of shear stresses in the anisotropy is significantly more pronounced (Figs.\ref{fig:D params nd}(b) and (d) ) and the directional diffusivity can change by a factor of two for different orientation. Regarding absolue values, curves in Fig. \ref{fig:D params nd}(c) and (e), show that uniaxial stresses do not have any significant effect on either the pre-exponential coefficient (D$^0$) or the effective energy barrier ($\Delta E$). However, shear stresses show some clear effects on both parameters; average D$^0$ increases almost linearly for stresses above 200 MPa, while $\Delta E$ also shows a linear increase above 200 MPa, with a small decrease below 200 MPa. 

The dependence of the diffusion tensor on elastic stress has been classically characterized by the strain-elasto-diffusion tensor, $\mathbb{d}^\epsilon$ \cite{Dederichs1978}. In this work, diffusion is defined as function of, defining  as stress-elasto-diffusion tensor, $\mathbb{D}^\sigma$, which under elastic strains is directly related with $\mathbb{d}^\epsilon$ through the elastic compliance, $\mathbb{S}$, as
\begin{equation}
    \mathbb{D}^\sigma = \frac{\partial \mathbf{D}}{\partial \boldsymbol{\sigma}} = \frac{\partial \mathbf{D}}{\partial \boldsymbol{\epsilon}}\frac{\partial \boldsymbol{\epsilon}}{\partial \boldsymbol{\sigma}} = \mathbb{d}^\epsilon : \mathbb{S}
    \label{eq:elasto-difu}
\end{equation}

To provide an analytical expression of the diffusivity tensor (Eq. \ref{eq:arrhenius2}) and its dependency with stress \ref{eq:elasto-difu} , the results obtained from simulating mobile hydrogen diffusivity for temperatures between 240 and 360 K and stress levels up to 800 MPa, presented in Figure \ref{fig:D params nd}, have been fit to a second order Taylor series expansion 
following Eq.\eqref{eq::D_taylor2}.

\begin{eqnarray}
D^0_{ij}= A_{ijkl} \sigma_{kl}^2 + B_{ijkl} \sigma_{kl} + C_{ij} \nonumber
\\ 
\label{eq::D_taylor2}
\Delta E = a_{kl} \sigma_{kl} + b
\end{eqnarray}
Considering the cubic symmetry of the undistorted $\alpha-\mathrm{Fe}$ lattice, the number of independent terms for the fourth rank tensors $A$ and $B$ reduced to six while in the case of the second rank tensors $C$ and $a$, only two independent terms are needed. 
The value of the parameters in Eqs. \eqref{eq::D_taylor2} and their symmetries are provided in \ref{eq::derivs_nd_sym}, together with the expressions to define the stress-elasto-diffusion tensor.

From this parametrization, it is interesting to note that both $D^0$ and $\Delta E$ are only weakly affected by uniaxial stresses, however, it is found that there is a high diffusivity dependence on shear stresses.
To illustrate this effect, the ratio of the diffusion coefficient between loaded and unloaded materials along directions parallel and perpendicular to the externally applied stress direction at different temperatures obtained by evaluating Eq.\eqref{eq:arrhenius2} is shown in Tables \ref{tab::uni_D_ratio} and \ref{tab::sh_D_ratio} for uniaxial and shear loading, respectively.

\begin{table}[ht!]
    \centering
    \caption{Ratio between directional diffusion coefficient between stressed and non-stressed material at different temperatures due to different levels of local uniaxial stress applied parallel ($\sigma^\parallel$) and perpendicularity ($\sigma^\perp$) to the diffusion direction}
    \label{tab::uni_D_ratio}
    \resizebox{1.0\textwidth}{!}{
    \begin{tabular}{|c||c|c|c|c|c|}
         \hline
        $\mathrm{D}\left(\sigma,T\right)/\mathrm{D}\left(\boldsymbol{0},T\right)$ & T = -100$^\circ \mathrm{C}$ & T = 25$^\circ \mathrm{C}$ & T = 200$^\circ \mathrm{C}$ & T = 500$^\circ \mathrm{C}$ & T = 800$^\circ \mathrm{C}$\\
         \hline
         \hline
         $\sigma^\parallel = 200$ MPa  & 1.03 & 1.02 & 1.01 & 1.01 & 1.01  \\
         \hline
         $\sigma^\parallel = 500$ MPa  & 1.06 & 1.03 & 1.02 & 1.01 & 1.00  \\
         \hline
         $\sigma^\parallel = 800$ MPa  & 1.07 & 1.02 & 1.00 & 0.98 & 0.98 \\
         \hline
         $\sigma^\perp = 200 $ MPa  & 1.03 & 1.02 & 1.01 & 1.01 & 1.01 \\
         \hline
         $\sigma^\perp = 500 $ MPa  & 1.10 & 1.07 & 1.06 & 1.05 & 1.04  \\
         \hline
         $\sigma^\perp = 800 $ MPa  & 1.20 & 1.15 & 1.12 & 1.11 & 1.10 \\
         \hline
    \end{tabular}}
\end{table}

\begin{table}[ht!]
    \centering
    \caption{Ratio between directional diffusion coefficient between stressed and non-stressed material at different temperatures due to different levels of local shear stress applied parallel ($\tau^\parallel$) and perpendicularity ($\tau^\perp$) to the diffusion direction}
    \label{tab::sh_D_ratio}
    \resizebox{1.0\textwidth}{!}{
    \begin{tabular}{|c||c|c|c|c|c|}
         \hline
         $\mathrm{D}\left(\tau,T \right)/\mathrm{D}\left(\boldsymbol{0},T\right)$ & T = -100$^\circ \mathrm{C}$ & T = 25$^\circ \mathrm{C}$ & T = 200$^\circ \mathrm{C}$ & T = 500$^\circ \mathrm{C}$ & T = 800$^\circ \mathrm{C}$\\
         \hline
         \hline
         $\tau^\parallel = 200 $ MPa & 1.22 & 1.47 & 1.61 & 1.71 & 1.76 \\
         \hline
         $\tau^\parallel = 500 $ MPa & 1.45 & 2.27 & 2.86 & 3.34 & 3.57 \\
         \hline
         $\tau^\parallel = 800 $ MPa & 1.40 & 2.89 & 4.14 & 5.34 & 5.95 \\
         \hline
         $\tau^\perp = 200 $ MPa & 0.78 & 0.94 & 1.03 & 1.10 & 1.13 \\
         \hline
         $\tau^\perp = 500 $ MPa & 0.74 & 1.17 & 1.47 & 1.71 & 1.83 \\
         \hline
         $\tau^\perp = 800 $ MPa & 0.71 & 1.45 & 2.11 & 2.69 & 3.00 \\
         \hline
    \end{tabular}}
\end{table}

Table \ref{tab::uni_D_ratio} shows the limited effect of uniaxial loading, with a maximum increase in the directional diffusivity of 20\%, when 800 MPa are applied at -100$^\circ \mathrm{C}$. This value is reduced to an increase of 16\% under these conditions when the mean diffusivity is considered ($\bar{D}=\frac{1}{3}(D^{\parallel}+2\, D^\perp)$. However, Table \ref{tab::sh_D_ratio} shows the significant effect caused by shear loading. In this case, the directional diffusivity becomes nearly six times higher (four times for the mean value) under the conditions of 800 MPa at 800$^\circ \mathrm{C}$. 

\subsubsection{Addition of dislocations}\label{sec:Dif_homo_dc}

The effect of the presence of dislocations has been evaluated by including a periodic array of screw dislocation dipoles in the $\left[ 1,1,1 \right]$ direction with the Burgers vector $\mathbf{b} =\pm \frac{a}{2}\left[ 1,1,1 \right]$ alternating sign along the plane with normal $\mathbf{n} = \left[ 0,1,\overline{1} \right]$, with a total dislocation density of $2 \times10^{14} \,\mathrm{m}/\mathrm{m}^3$ (Fig. \ref{fig:scheme}). Dislocations are represented by phase fields, as described in Section \ref{sec:Disloc}. A new set of OkMC simulations were performed, including the presence of the dislocations, sharing the same conditions used in the previous section (number of point defects, cell size, etc.). Simulations were performed for the different levels of uniaxial and shear stresses used without including dislocations.

In the absence of external stress, the local elastic distortion introduced by the presence of dislocations modifies the effective diffusion coefficient and breaks the lattice cubic symmetry, 
introducing for example non-hydrostatic terms in the diffusivity that were zero for a pristine cubic lattice. When external stress is introduced, this loss of symmetry due to the presence of dislocations causes the number of independent parameters defining the dependencies of diffusivity on the stress (Eq. \ref{eq::D_taylor2}) also to increase. 

\begin{figure}[p!]
    \centering
    \begin{subfigure}[b]{0.95\textwidth}
         \centering
         \includegraphics[width=\textwidth]{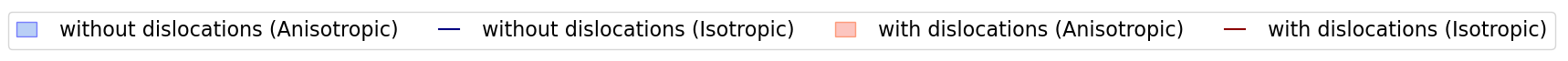}
    \end{subfigure}
    
    \begin{adjustbox}{max totalsize={\textwidth}{0.27\textheight}}
    \begin{subfigure}[b]{0.50\textwidth}
         \centering
         \includegraphics[width=\textwidth]{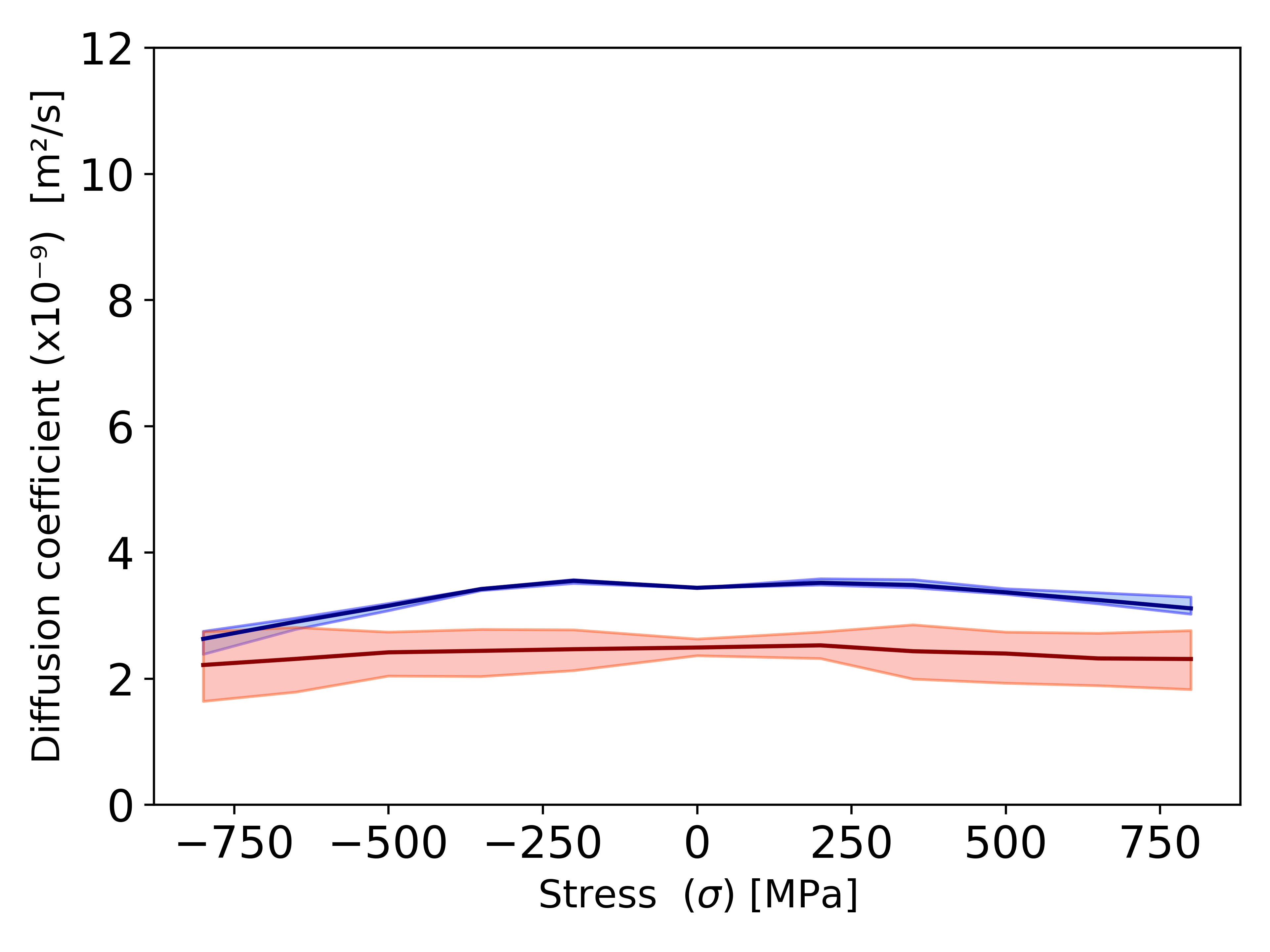}
         \subcaption[]{}
    \end{subfigure}
    \begin{subfigure}[b]{0.50\textwidth}
         \centering
         \includegraphics[width=\textwidth]{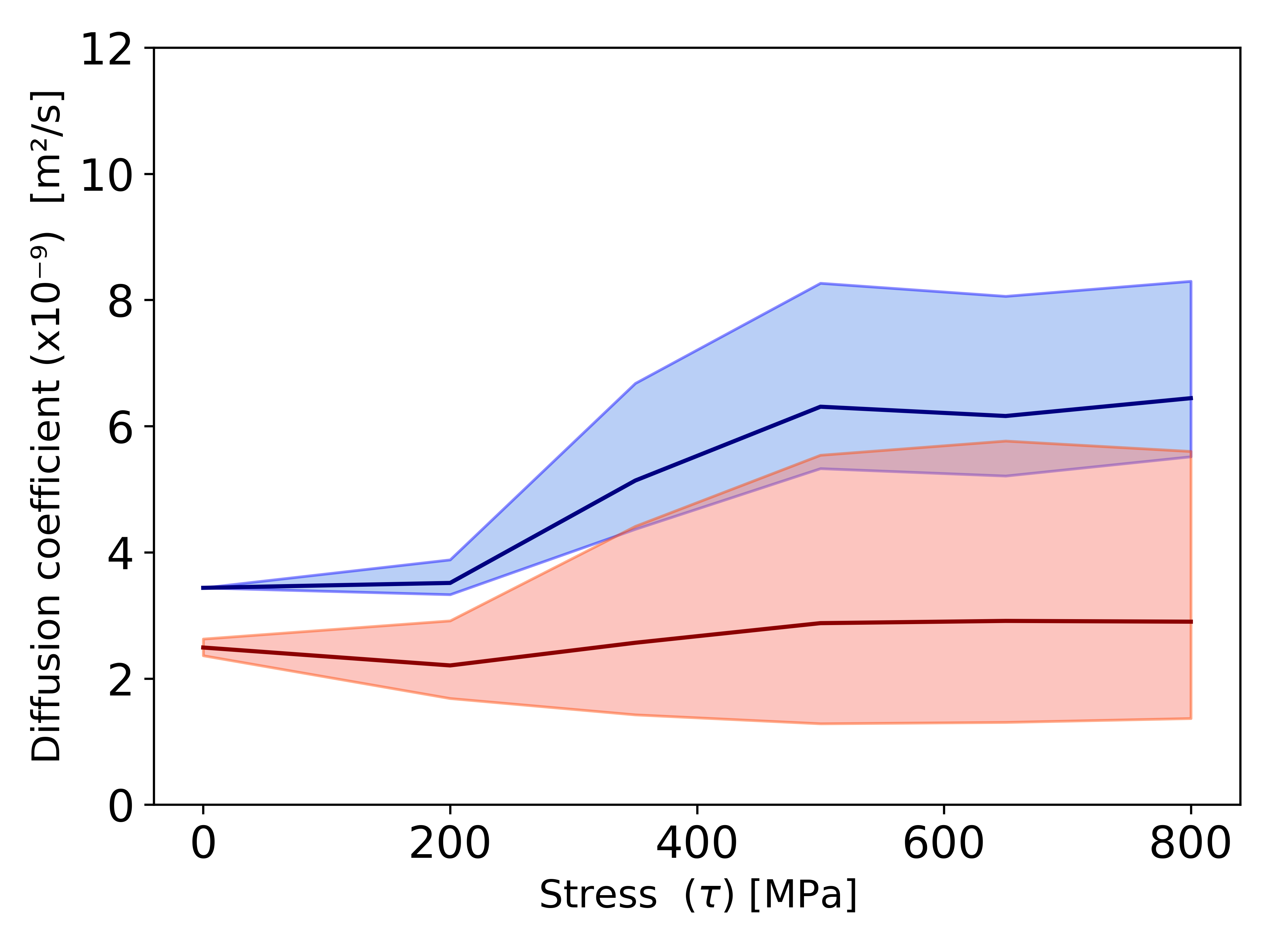}
         \subcaption[]{}
    \end{subfigure}
    \end{adjustbox}
    \begin{adjustbox}{max totalsize={\textwidth}{0.27\textheight}}
    \begin{subfigure}[b]{0.50\textwidth}
         \centering
         \includegraphics[width=\textwidth]{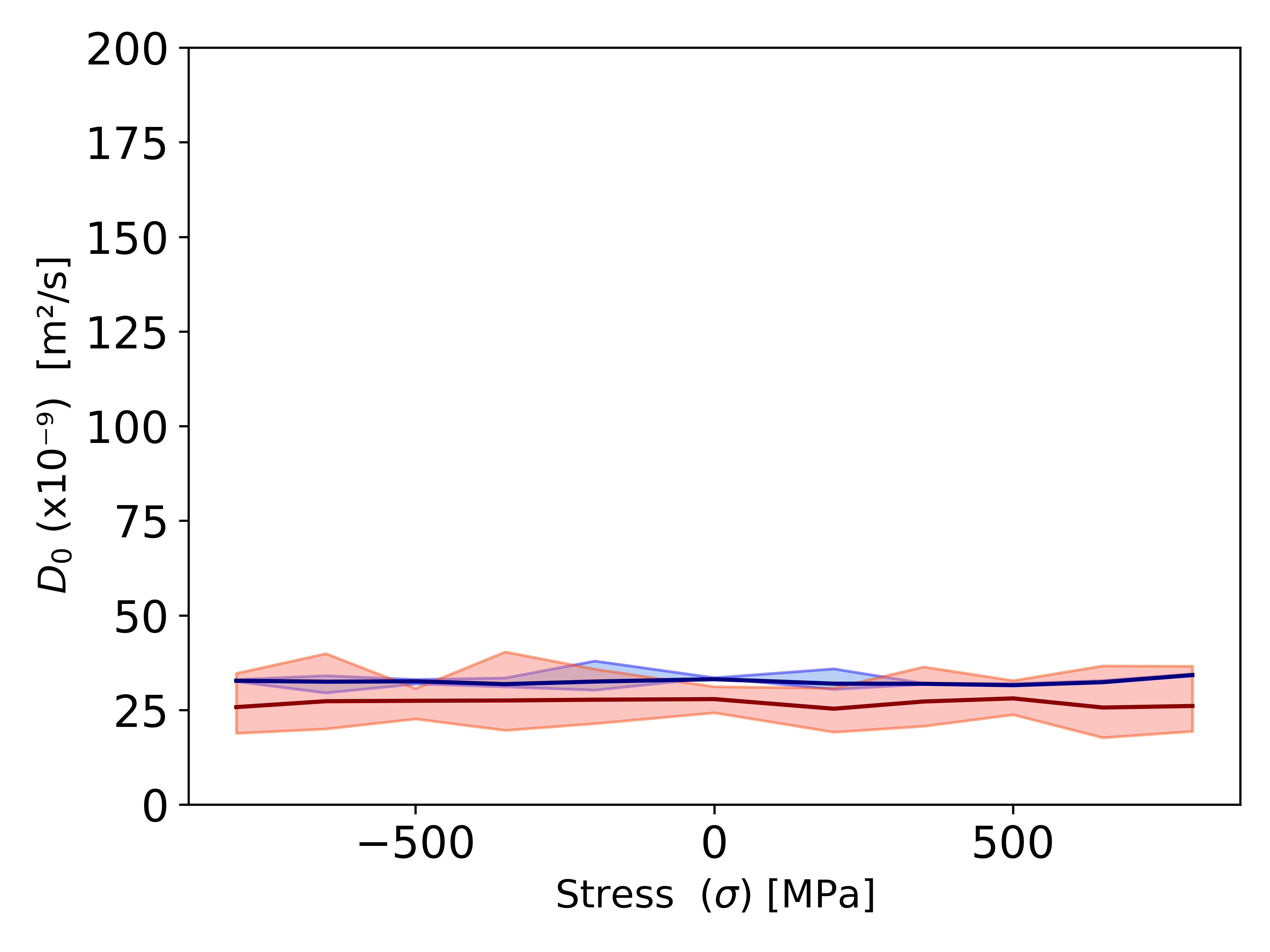}
         \subcaption[]{}
    \end{subfigure}
    \begin{subfigure}[b]{0.50\textwidth}
         \centering
         \includegraphics[width=\textwidth]{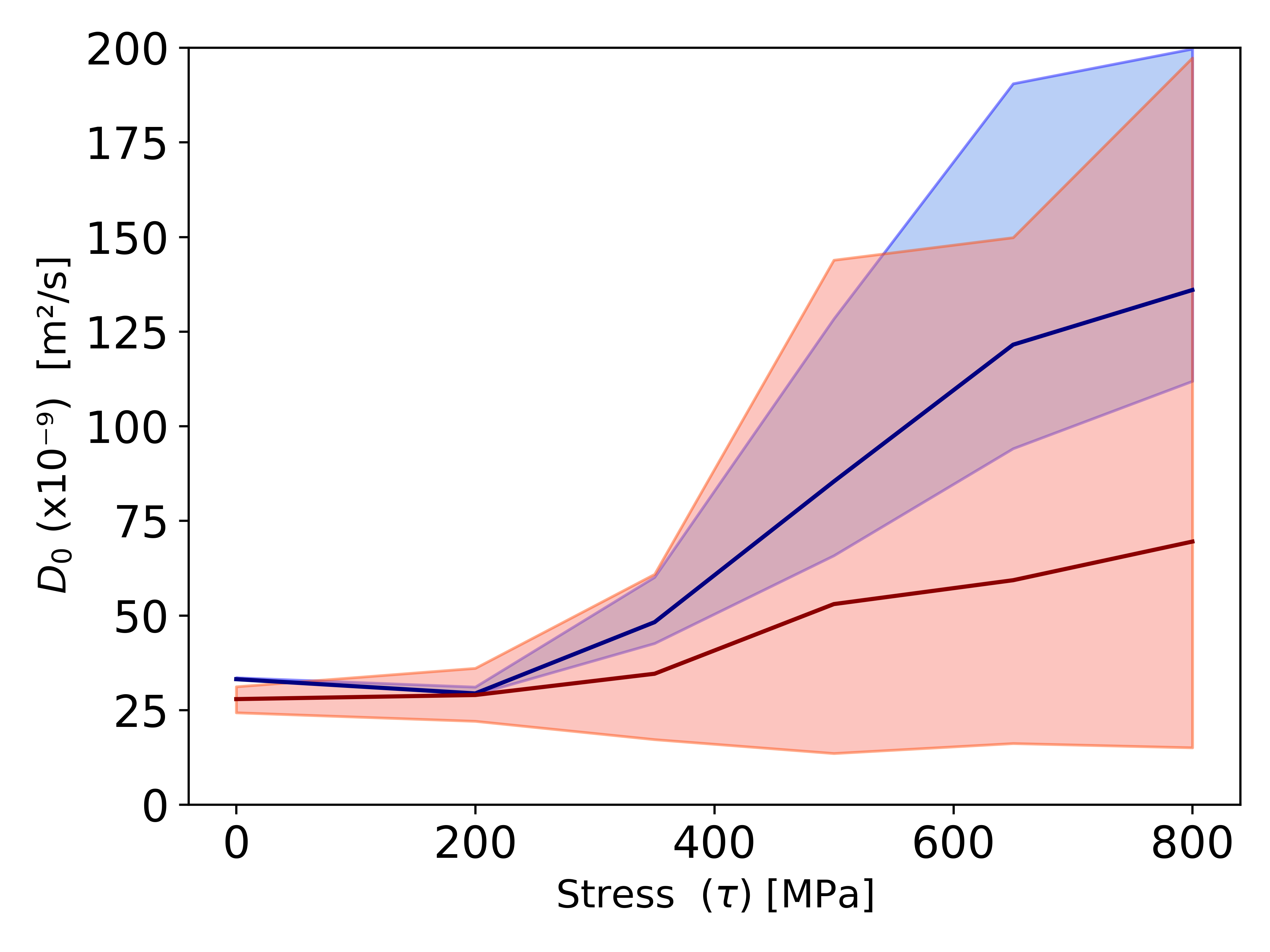}
         \subcaption[]{}
    \end{subfigure}
    \end{adjustbox}
    \begin{adjustbox}{max totalsize={\textwidth}{0.27\textheight}}
    \begin{subfigure}[b]{0.50\textwidth}
         \centering
         \includegraphics[width=\textwidth]{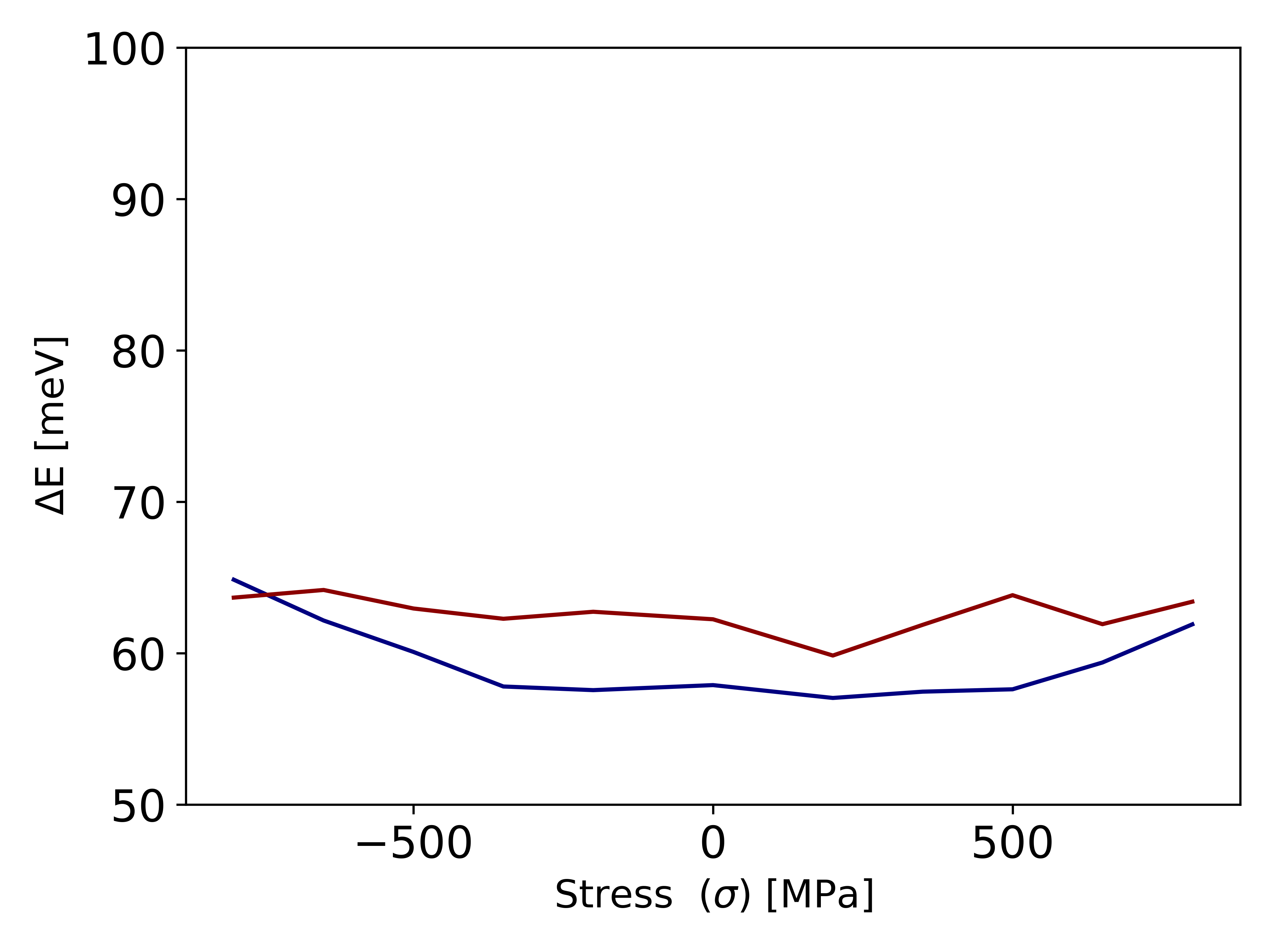}
         \subcaption[]{}
    \end{subfigure}
    \begin{subfigure}[b]{0.50\textwidth}
         \centering
         \includegraphics[width=\textwidth]{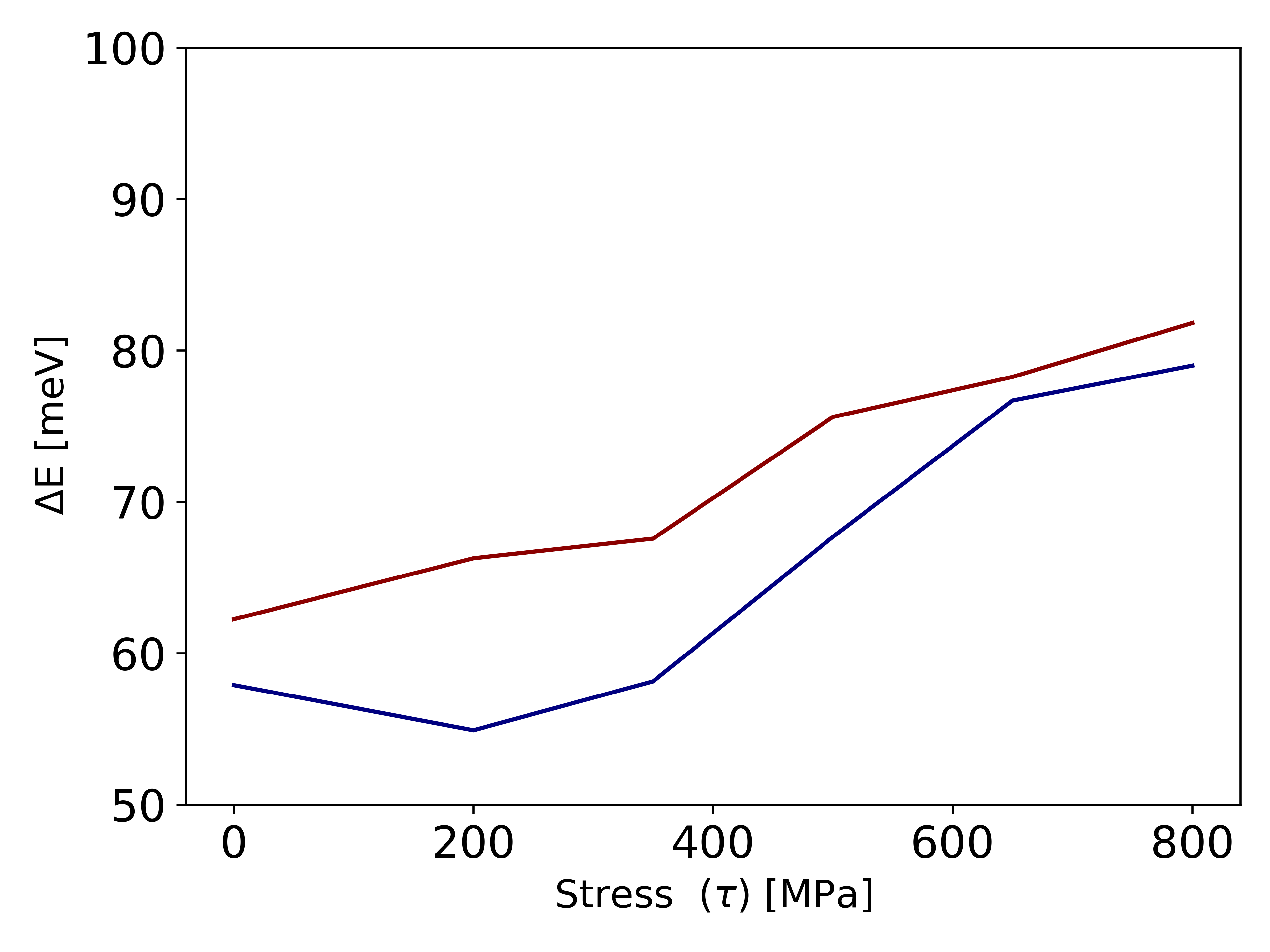}
         \subcaption[]{}
    \end{subfigure}
    \end{adjustbox}
    
     \caption{Effect of stress in the diffusivity. Diffusion coefficient at 300 K (D) (a,b) pre-exponential coefficient (D$^0$) (c,d) and effective energy barrier ($\Delta E$) (e,f). Under uniaxial stress (a,c,e) and shear stress (b,d,f). Solid lines represent average parameters, shadows represent the region occupied by the directional counterparts.
     }
     \label{fig:D parmas d}
\end{figure}

The effect of stress on the diffusivity parameters is shown in Fig.\ref{fig:D parmas d}, where the results obtained including dislocations are plotted together with the results of the crystal without dislocations. The values of the parameters defined in Eq.\eqref{eq::D_taylor2} for the case with dislocation are listed in 
\ref{eq::derivs_d}.
The curves in Fig.\ref{fig:D parmas d} include with lines the average diffusion coefficient at 300 K, the prefactor and the effective barrier as a function of the applied uniaxial or shear stress, and in shadow regions the maximum and minimum values of the directional diffusivity and corresponding prefactors and energy barriers.

A general observation of Fig. \ref{fig:D parmas d} shows that the presence of a dislocation causes a decrease in the diffusivity for all the stress states considered (see Figs. \ref{fig:D parmas d}(a) and (b) ). The effect of the stress found is similar to the pristine case, uniaxial stress has a much lower effect than shear. Regarding the anisotropy, it is observed that for zero stress in the absence of dislocations, the diffusivity and the Arrhenius prefactor are isotropic (Fig. \ref{fig:D params nd}(a,b,c,d), single value of directional parameters). In contrast, in the presence of dislocations these parameters present anisotropy, which is observed by the shadowed red area in the Fig.\ref{fig:D params nd}(a,b,c,d). 

To quantify the impact of the presence of dislocations in the diffusivity of hydrogen in $\alpha-\mathrm{Fe}$, Table \ref{tab:d_D_ratio} includes the ratios of the mean and directional diffusion coefficients between dislocated and pristine lattices for different temperatures for an unstressed crystal. In addition, a detailed list of the difference between the parameters defining the diffusion tensor and energy barrier (Eq. \eqref{eq::D_taylor2}) between a material with and without dislocations is presented in \ref{eq::derivs_d_delta_sym}.
The differences, even at room temperature, reach 30\% of a reduction in diffusivity. This reduction is due to the presence of shear stresses around the dislocation, which on average decreases the frequency of jumps in hydrogen. This change should not be confused with the trapping of hydrogen in the dislocation line. Here, it is assumed that the hydrogen in the dislocation core is in equilibrium, and therefore diffusivity only accounts for non-trapped hydrogen.

\begin{table}[ht!]
    \centering
    \caption{Ratio of mean, $\bar{D}$, and directional, $D^{u}$, diffusion coefficient due to the presence of a $\rho =2\cdot 10 ^{14} \mathrm{m}/\mathrm{m}^3$ $\frac{1}{2}\left[ 111 \right] \left( 110 \right)$ dislocation density at different temperatures.}
    \label{tab:d_D_ratio}
    \begin{singlespace}
    \resizebox{1.0\textwidth}{!}{
    \begin{tabular}{|c||c|c|c|c|c|c|c|c|}
        \hline
        $D^\mathbf{u}(\rho,T)/D^\mathbf{u}(0,T)$ & T = $-100^\circ \mathrm{C}$ & T = $25^\circ \mathrm{C}$ & T = $200^\circ \mathrm{C}$ & T = $500^\circ \mathrm{C}$ & T = $800^\circ \mathrm{C}$ & T $\rightarrow\infty$ ($D^0$) \\
        \hline
        \hline
        $\mathbf{u} = \mathrm{isotropic}$ & 0.58 & 0.7 & 0.77 & 0.82 & 0.84 & 0.90 \\
        \hline
        $\mathbf{u} = \left[1\,0\,0\right]$ & 0.61 & 0.72 & 0.79 & 0.83 & 0.85 & 0.91 \\
        \hline
        $\mathbf{u} = \left[0\,1\,0\right]$ & 0.60 & 0.72 & 0.79 & 0.84 & 0.86 & 0.93 \\
        \hline
        $\mathbf{u} = \left[0\,0\,1\right]$ & 0.53 & 0.65 & 0.73 & 0.78 & 0.80 & 0.87 \\
        \hline
        $\mathbf{u} = \left[1\,1\,1\right]$ & 0.56 & 0.68 & 0.75 & 0.79 & 0.82 & 0.88 \\
        \hline
        $\mathbf{u} = \left[\bar{1}\,1\,0\right]$ & 0.64 & 0.76 & 0.83 & 0.88 & 0.90 & 0.96 \\
        \hline
        $\mathbf{u} = \left[2\,\bar{1}\,\bar{1}\right]$ & 0.62 & 0.74 & 0.81 & 0.86 & 0.88 & 0.94 \\
        \hline
        
    \end{tabular}}
    \end{singlespace}
\end{table}

From the results in Table \ref{tab:d_D_ratio} and the full set of results in the Appendix, the following conclusions on the effect of the presence of dislocations can be summarized.
\begin{itemize}
    \item The presence of dislocations induces anisotropy in the diffusivity tensor $\mathbf{D}$ independently of the external stress applied.
    \item In the absence of external stresses:
    \begin{enumerate}
        \item The presence of dislocations reduces diffusivity by two different mechanisms: reduction of the pre-exponential coefficient $\mathrm{D}^0$, and increase in the effective diffusion barrier in the material.
        \item There is a consistent reduction in the diffusivity along all crystalline directions.
        Considering the error introduced due to the statistical sampling, the maximum directional diffusivity is observed along the direction perpendicular to the shear plane, $\bar{\mathrm{n}}=\left[ 0 1 \bar1 \right]$, while the dislocation direction, $\bar{\mathrm{t}}=\left[ 1 1 1 \right]$, exhibits the minimum directional diffusivity.
        \item The increase in temperature reduces the effect of dislocations on the diffusivity of hydrogen.
    \end{enumerate}
    \item The effect of external stresses, 
    \begin{enumerate}
        \item The effective energy barriers with diffusion are increased slightly ($\approx6$meV), modifying the temperature dependence on the diffusivity tensor.
        \item There is a consistent reduction in the effect of external stresses on $\mathbf{D}^0$, both in linear and quadratic dependence. Significantly reduces the effect of external loading in hydrogen diffusivity.
    \end{enumerate}
\end{itemize}

\subsection{Distribution of mobile hydrogen}\label{sec::mobH_dc}

The presence of dislocations introduces new traps for hydrogen. It is well known that the depths of the hydrogen potential wells in the dislocation core are very deep \cite{Itakura2012,Itakura2013}, so once the core is saturated, trapped hydrogen does not contribute to the diffusion process. Moreover, it has been reported that the time required for hydrogen to be trapped in the dislocation core is very small, because this process is instantaneous compared to the time required for diffusion \cite{Matsumoto2022}.

Being said that, once the dislocation cores are saturated, the presence of dislocations still plays an important role in hydrogen diffusion, as has been shown in previous sections. In particular, the stress fields around the dislocations change the energy barriers and in consequence the jump frequencies. From a macroscopic view point, this effect leads to a reduction in the effective diffusivity (Tables \ref{tab::uni_D_ratio} and \ref{tab::sh_D_ratio}). On the microscale, this effect implies a heterogeneous distribution of hydrogen in the material, resulting in dependence on the position of the average volumetric residence time of hydrogen.

In order to evaluate this effect, the mobile hydrogen residence time around dislocations has been studied. Two sets of OKMC simulations with and without dislocations were evaluated. Each simulation had a 132$^3$ nanometer box, a simulated time of three nanoseconds and three random seeds were evaluated for the initial hydrogen distribution for each set.  A hydrogen atomic concentration of $3 \times 10^{-6}$, and an iron vacancy concentration of $1 \times 10^{-7}$ have been used.
In order to evaluate the spatial distribution of residence time of H the simulation box was discretized into a $(5\mathrm{a}_{\alpha-Fe})^3$ regular voxel grid, and the the volumetric residence time, $\tau_V^H$, was computed as the time spent the individual particles in each voxel in the latter half of the simulation Eq.\eqref{eq:vol_res_time}. 

\begin{equation}\label{eq:vol_res_time}
    \tau_V^H(\mathbf{x})=\int_{t_f/2}^{t_f} c_H (\mathbf{x}) dt
\end{equation}

The areal residence time ,$\tau_A^H$, corresponds to the prowas evaluated by integrating the volumetric residence time, $\tau_V^H$, along the direction $\left[ 1 1 1 \right]$ (screw dislocation), considering the periodicity of the simulation cell.

\begin{figure}[ht]
    \centering  
    \begin{subfigure}[br]{0.49\textwidth}
    \begin{adjustbox}{max width=\textwidth}
        \includegraphics{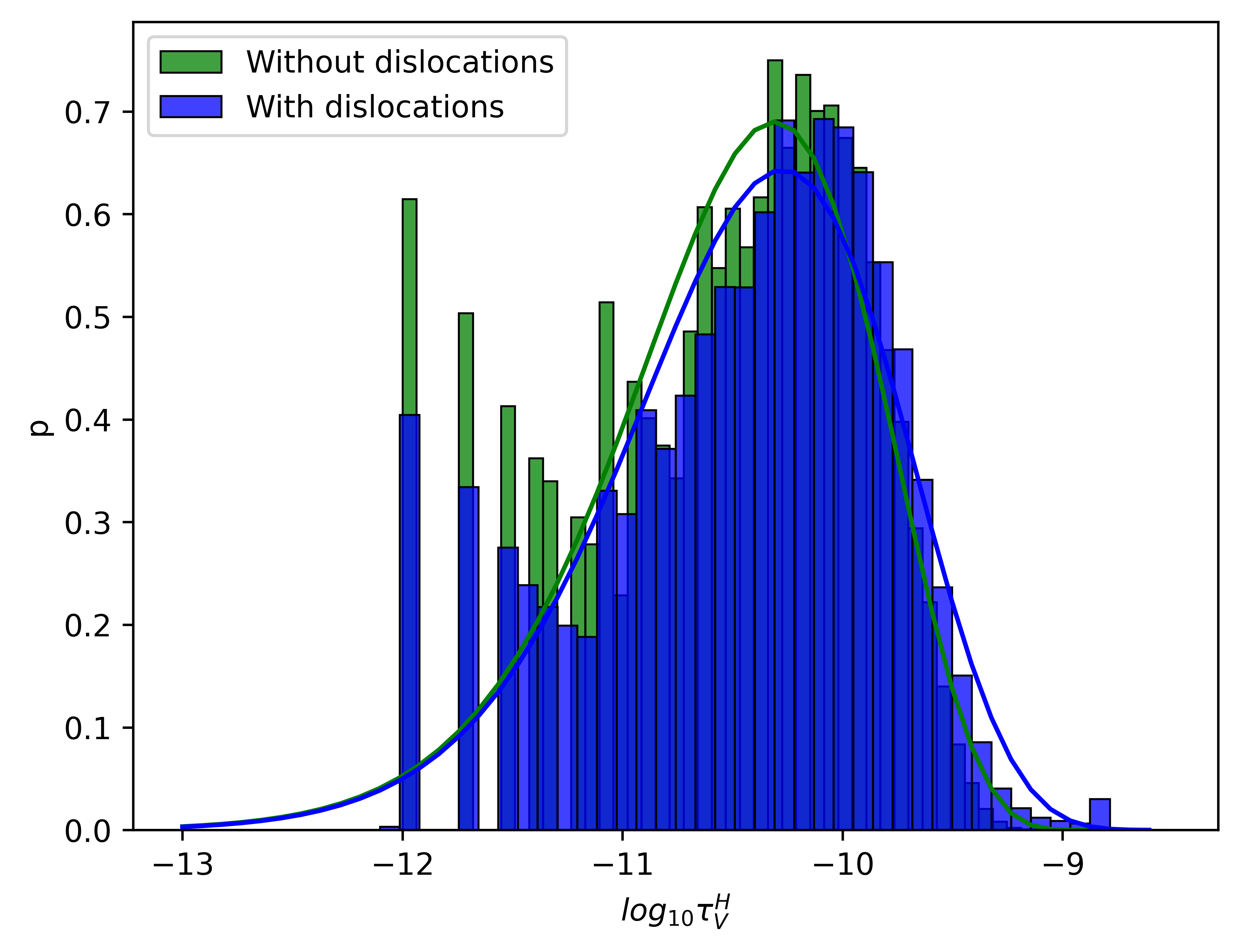}
    \end{adjustbox}
    \subcaption[]{}
    \end{subfigure}
    \begin{subfigure}[bl]{0.49\textwidth}
    \begin{adjustbox}{max width=\textwidth}
        \includegraphics{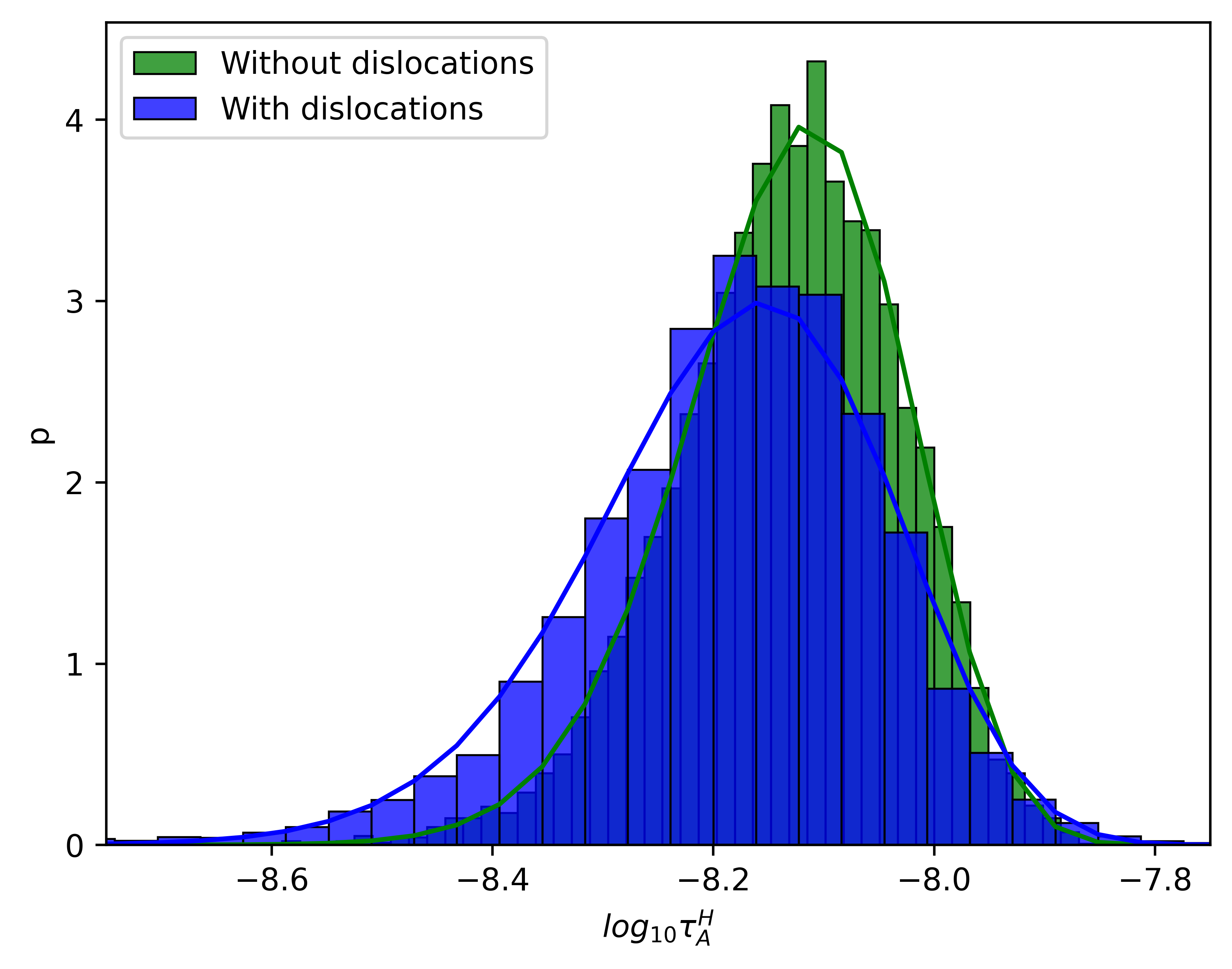}
    \end{adjustbox}
    \subcaption[]{}
    \end{subfigure}
    \caption{Probability distributions of residence time of mobile hydrogen with and without dislocations: a) volumetric distribution b) areal distribution}
    \label{fig::mobH_restime_dc}
\end{figure}

From these simulations, the following information could be obtained:
\begin{itemize}
    \item The H residence time distributions were qualitatively identical for different OkMC runs, even when starting with different random H distributions, but the results in pristine material and in the presence of dislocations presented strong differences.
    \item The distributions of the areal and volumetric hydrogen residence times were best fitted by Gamma distributions (Eq.\eqref{eq::Gamma_dist}) of their logarithm (Fig.\ref{fig::mobH_restime_dc})
    \begin{equation} \label{eq::Gamma_dist}
     \Gamma(x; k,\theta,x_0) = (x-x_0)^{k-1} \frac{e^{\frac{-(x-x_0)}{\theta}}}{\theta^{k}\Gamma(k)}
    \end{equation}
where $x$ here is the decimal logarithm of the residence time. The parameters for $(k $, $\theta, x_0)$ for volumetric and areal distributions in the cases with and without dislocations are given in Table \ref{tab:gamma}
    
    \item The addition of dislocations modifies the lattice residence time of hydrogen 
    by reducing the median areal residence time of hydrogen, $\tau_A^H$. This reduction is coupled with an increase in the probability density for longer residence times. 
\end{itemize}

\begin{table}[ht!]
    \centering
    \caption{Parametrization of the gamma functions describing hydrogen areal and volumetric residence time in a simulation cell with and without dislocations.}
    \label{tab:gamma}
    \begin{singlespace}
    \resizebox{0.6\textwidth}{!}{
    \begin{tabular}{|c||c|c|c|}
        \hline
        $\Gamma(x; k,\theta,x_0)$ & $k$ & $\theta$& $x_0$\\
        \hline
        \hline
        $\tau_V^H$ without dislocations & 0.6168 & 0.217 & 10.53 \\
        \hline
        $\tau_V^H$ with dislocations  & 0.6450 & 0.188 & 10.46\\
        \hline
        $\tau_A^H$ without dislocations  & 0.1022 & 0.021 & 8.13 \\
        \hline
        $\tau_A^H$ with dislocations  & 0.1376 & 0.028 & 8.18 \\
        \hline        
    \end{tabular}}
    \end{singlespace}
\end{table}

\begin{figure}[ht]
    \centering
    \begin{subfigure}{1\textwidth}
        \includegraphics[scale=0.4]{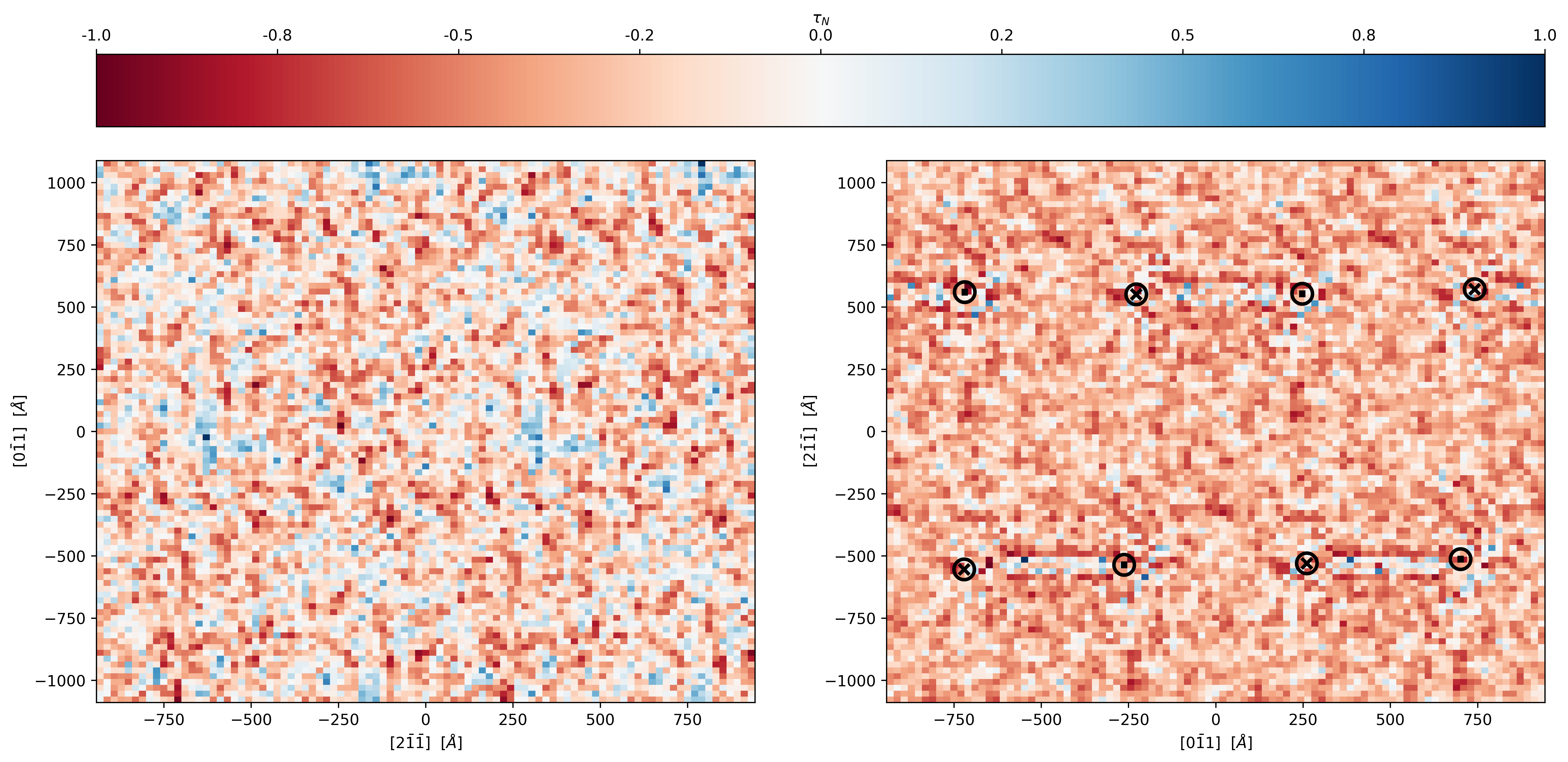}
    \end{subfigure}
    \caption{Normalized residence time of mobile hydrogen in the $\left( 1 1 1 \right)$ plane: a) without dislocations b) with dislocations}
    \label{fig:mobH_dc}
\end{figure}

In order to understand the effect of dislocations on the hydrogen residence time distribution in space, the normalized value of the areal residence time
along the $\left( 1 1 1 \right)$ plane, $\tau_N$ (Eq.\eqref{eq::tauN}), has been represented with colors in Fig.\ref{fig:mobH_dc}).

\begin{equation}\label{eq::tauN}
    \log_{10} \tau_N = \frac{\log_{10}(\tau_A^H)-\left< \log_{10}(\tau_A^H) \right>} {\max \left( \max \left( \log_{10}(\tau_A^H)\right)-\left< \log_{10}(\tau_A^H) \right>, \left< \log_{10}(\tau_A^H) \right> - \min \left( \log_{10}(\tau_A^H)\right)\right)}
\end{equation}

Figure \ref{fig:mobH_dc} shows how the addition of dislocations changes the residence time of hydrogen in the volume between dislocation dipoles, with an above-average residence time volume, surrounded by a depletion zone surrounding the sheared plane.

\section{Conclusions}\label{sec:Conclusions}

The effect of local stress state on the diffusivity and  distribution of absorbed interstitial hydrogen in body-centered cubic iron has been studied.

To this aim, a multiscale model based on a lattice object kinetic Monte Carlo code (OkMC) has been developed. The OkMC model drives the movement of a hydrogen and vacancies within the lattice, and has been particularized for BCC iron. At the nano-level, the model includes the energies of H and vacancies in different positions of the lattice, obtained by DFT simulations. Moreover, the effect of stress on this chemical energy has also been obtained from DFT. At the microscopic level, the effect of the elastic strains caused by different defects is also accounted in the model by modifying the energy barriers, which allows to reproduce preferential diffusion paths depending on the strain gradients. In the model, dislocations are included and represented by phase-fields to reproduce adequately their elastic strains in a non-isotropic matrix as well as the core regions.


Using the code developed the diffusion tensor in the presence of simple macroscopic homogeneous elastic fields, in pristine lattice and in the presence of dislocations, has been evaluated. The main conclusions are
\begin{itemize}
    \item The external stress has a strong influence in the diffusivity, specially shear stresses. In this case, even at room temperature diffusivity is increased by a factor of two for 500MPa. The increase can reach up to a factor of 6 for high temperature (800$^\circ$C) and shears of 800MPa.
\item The diffusion tensor, which in a relaxed configuration is isotropic, develops anisotropy when shear stresses are applied. The difference between diffusivities in the sheared plane and perpendicular to it can reach a factor of 2
\item The presence of dislocations strongly affects the diffusivity. First, it breaks its isotropy even in the absence of stress. Second, it reduces its value by the effect of the local stresses around the dislocations (30\%  of reduction in the absence of external stress fields at room temperature)
\end{itemize}

Finally, the results of OkMC simulations have been used to provide closed expression of the diffusivity tensor of hydrogen in the presence of dislocations under different stress and temperature conditions has been provided. This expression can be used in higher-scale models to reproduce proper hydrogen diffusion in BCC-Fe without the requirement to consider individual hydrogen atoms, allowing the faithful simulation of larger length and time scales.

\section{Acknowledgements}
The authors thank the Ministry of Science and Innovation of Spain for the financial support from Grant TED2021-130255B-C32, funded by 
MCIN/AEI /10.13039/501100011033.

G. Álvarez thanks the Scholarship provided by the Universidad Politécnica de Madrid and the Ministry for Universities (Ref: FPU20/05495).

The authors gratefully acknowledge the Universidad Politécnica de Madrid for providing computing resources on the Magerit Supercomputer. 

\bibliographystyle{elsarticle-num} 
\bibliography{references}

\appendix

\newpage

\setcounter{page}{1}
\setcounter{table}{0}
\setcounter{equation}{0}
\section{Voigt/Nye notation for tensors}
The numerical parametrization of the parameters required to evaluate the diffusivity tensor at arbitrary states of local stress and temperature are presented in this appendix following Voigt/Nye notation (Eq.\eqref{eq:Voigt}) taking advantage of the minor symmetries present in all the relevant tensors.

\begin{eqnarray}\label{eq:Voigt}
    \mathbb{T}_{ijkl}\rightarrow\mathbb{T}_{\alpha \beta} \\
    \mathbf{T}_{ij}\rightarrow\mathbf{T}_{\alpha}
\end{eqnarray}

\begin{eqnarray}\label{eq:Voigt2}
    \{i,j\}=\{1,1\} \leftrightarrow \alpha=1 \\
    \{i,j\}=\{2,2\} \leftrightarrow \alpha=2 \\
    \{i,j\}=\{3,3\} \leftrightarrow \alpha=3 \\
    \{i,j\} \in \left\{\{2,3\}, \{3,2\}\right\}\leftrightarrow \alpha=4 \\
    \{i,j\} \in \left\{\{1,3\}, \{3,1\}\right\} \leftrightarrow \alpha=5 \\
    \{i,j\} \in \left\{\{1,2\}, \{2,1\}\right\}\leftrightarrow \alpha=6 
\end{eqnarray}

\section{Numerical evaluation of elastodiffusion tensors}\label{ap:elastodif}
The value of the parameters in the Taylor expansion of the diffusion tensor and the energy barrier:
\begin{eqnarray*}
D^0_{ij}= A_{ijkl} \sigma_{kl}^2 + B_{ijkl} \sigma_{kl} + C_{ij} \nonumber
\\ 
\Delta E = a_{kl} \sigma_{kl} + b
\end{eqnarray*}
for the pristine material ($\mathbf{D}^0, \; \Delta E^0$), the material with dislocations ($\mathbf{D}^\mathrm{d}, \; \Delta^2 E$), and their difference ($\Delta\mathbf{D}, \; \Delta E^\mathrm{d}$) are presented in this appendix.

This parameterization allows the determination of the stress-elastodiffusivity tensor $\mathbb{D}^\sigma$ using Eq.\eqref{eq::D_ABCab}

\begin{equation}\label{eq::D_ABCab}
\mathbb{D}^\sigma_{ijkl} (\boldsymbol{\sigma},T) = \left(
2 A_{ijkl} \sigma_{kl} + B_{ijkl}   -
(A_{ijkl} \sigma_{kl}^2 + B_{ijkl} \sigma_{kl} + C_{ij})   \frac{a_{kl}}{k_BT}   \right)
exp\left(\frac{-(a_{kl} \sigma_{kl} + b)}{k_BT}\right)
\end{equation}

\subsection{$\mathbf{D}^0, \; \Delta E^0$}\label{eq::derivs_nd_sym}
\begin{equation*}
\left.
\begin{aligned} 
&A^0_{ijkl}=10^{-21} \times
\begin{bmatrix}
 -3,02  & 4,01  & 4,01  & 203,38  & 130,59  & 130,59 \\
  4,01  &-3,02  & 4,01  & 130,59  & 203,38  & 130,59 \\
  4,01  & 4,01  & -3,02 & 130,59  & 130,59  & 203,38 \\
  -0,04 &  0,40 &  0,40 & -60,23  &  12,98  &  12,98 \\
   0,40 & -0,04 &  0,40 &  12,98  & -60,23  &  12,98 \\
   0,40 & 0,40  & -0,04 &  12,98  &  12,98  & -60,23
\end{bmatrix}
\mathrm{m}^2\mathrm{s}^{-1}\mathrm{Pa}^{-2}\\
\end{aligned} 
\right.
\end{equation*}
\begin{equation*}
\left.
\begin{aligned} 
&B^0_{ijkl}=10^{-18} \times
\begin{bmatrix}
  1.03  & -0,13  & -0,13 &  89,47  &  4,32  & 4,32 \\
 -0,13  &  1.03  & -0,13 &  4,32   & 89,47  & 4,32  \\
 -0,13  & -0,13  &  1.03 &  4,32   &  4,32  & 89,47  \\
 -1,22  & -0,23  & -0,23 &  56,67  & 92,81  & 92,81 \\
 -0,23  & -1,22  & -0,23 &  92,81  & 56,67  & 92,81  \\
 -0,23  & -0,23  & -1,22 &  92,81  & 92,81  & 56,67  
\end{bmatrix}
\mathrm{m}^2\mathrm{s}^{-1}\mathrm{Pa}^{-1}\\
\end{aligned} 
\right.
\end{equation*}
\begin{equation*}
\left.
\begin{aligned} 
&C^0_{ij}=
10^{-9} \times
\begin{bmatrix}
  29.44 & 29.44 & 29.44 & -2.34 & -2.34 & -2.34
\end{bmatrix}
\mathrm{m}^2\mathrm{s}^{-1}\\
\end{aligned} 
\right.
\end{equation*}
\begin{equation*}
\left.
\begin{aligned} 
&a^0_{kl}=
10^{-12} \times
\begin{bmatrix}
  -1.91 & -1.91 & -1.91 & 32.12 & 32.12 & 32.12
\end{bmatrix}
\mathrm{eV \,Pa}^{-1}\\
\end{aligned} 
\right.
\end{equation*}
\begin{equation*}
\left.
\begin{aligned} 
&b^0= 
10^{-3} \times 55.89
\mathrm{eV}
\end{aligned} 
\right.
\end{equation*}

\subsection{$\mathbf{D}^\mathrm{d}, \; \Delta E^\mathrm{d}$}\label{eq::derivs_d}
\begin{equation*}
\left.
\begin{aligned} 
&A^{\mathrm{d}}_{ijkl}=10^{-21} \times
\begin{bmatrix}
  0.95  & 5.32   & -3.88  & 7.26   & -29.32  & -20.59 \\
  -5.12 & -4.60  & -10.12 & -2.17  & 143.92  & 75.50  \\
  0.63  & 3.64   & 2.11   & 30.29  & 124.80  & 167.00 \\
  -4.33 & 3.71   & 3.08   & -14.07 & 52.20   & 98.35  \\
  1.03  & 10.54  & -1.79  & 34.20  & -57.56  & -37.36 \\
  -3.49 & 5.87   & -2.59  & -39.33 & -25.02  & -31.57
\end{bmatrix}
\mathrm{m}^2\mathrm{s}^{-1}\mathrm{Pa}^{-2}\\
\end{aligned} 
\right.
\end{equation*}
\begin{equation*}
\left.
\begin{aligned} 
&B^{\mathrm{d}}_{ijkl}=10^{-18} \times
\begin{bmatrix}
  2.26  & 0.11   & 0.46  & 115.50  & 7.17   & -1.42  \\
  -0.34 & -0.64  & 2.33  & -27.22  & -1.16  & 48.71  \\
  1.66  & -4.56  & -0.09 & 22.08   & 25.56  & -0.65  \\
  0.64  & 0.85   & -2.39 & 2.04    & 44.82  & 23.03  \\
  -0.64 & -0.03  & 0.49  & 53.07   & 63.77  & 41.35  \\
  0.58  & 2.58   & 0.09  & 31.11   & 23.45  & 27.58  
\end{bmatrix}
\mathrm{m}^2\mathrm{s}^{-1}\mathrm{Pa}^{-1}\\
\end{aligned} 
\right.
\end{equation*}
\begin{equation*}
\left.
\begin{aligned} 
&C^{\mathrm{d}}_{ij}=
10^{-9} \times
\begin{bmatrix}
  26.79 & 27.27 & 25.63 & -0.06 & -1.29 & 0.31 
\end{bmatrix}
\mathrm{m}^2\mathrm{s}^{-1}\\
\end{aligned} 
\right.
\end{equation*}
\begin{equation*}
\left.
\begin{aligned} 
&a^{\mathrm{d}}_{kl}=
10^{-12} \times
\begin{bmatrix}
  1.28 & 2.38 & -4.39 & 22.98 & 24.10 & 29.77
\end{bmatrix}
\mathrm{eV \,Pa}^{-1}\\
\end{aligned} 
\right.
\end{equation*}
\begin{equation*}
\left.
\begin{aligned} 
&b^{\mathrm{d}}= 
10^{-3} \times 61.74
\mathrm{eV}
\end{aligned} 
\right.
\end{equation*}

\subsection{$\Delta\mathbf{D}=\mathbf{D}^\mathrm{d}-\mathbf{D}^\mathrm{0}, \; \Delta ^2 E = \Delta E^\mathrm{d} - \Delta E^\mathrm{0}$}\label{eq::derivs_d_delta_sym}

\begin{equation*}
\left.
\begin{aligned} 
& \Delta A_{ijkl}=10^{-21} \times
\begin{bmatrix}
3.97 & 1.31 & -7.89 & -196.12 & -159.91 & -151.18 \\
-9.13 & -1.58 & -14.13 & -132.76 & -59.46 & -55.08 \\
-3.38 & -0.37 & 5.13 & -100.3 & -5.79 & -36.38 \\
-4.29 & 3.31 & 2.68 & 46.16 & 39.22 & 85.37 \\
0.63 & 10.58 & -2.19 & 21.22 & 2.67 & -50.34 \\
-3.89 & 5.47 & -2.55 & -52.3 & -38.0 & 28.66 \\
\end{bmatrix}
\mathrm{m}^2\mathrm{s}^{-1}\mathrm{Pa}^{-2}\\
\end{aligned} 
\right.
\end{equation*}
\begin{equation*}
\left.
\begin{aligned} 
& \Delta B_{ijkl}=10^{-18} \times
\begin{bmatrix}
1.22 & 0.24 & 0.59 & 26.03 & 2.85 & -18.5 \\
-0.21 & -1.67 & 2.46 & -31.55 & -101.02 & 44.39 \\
1.79 & -4.43 & -1.12 & 17.76 & 21.24 & -95.95 \\
1.86 & 1.08 & -2.16 & -54.63 & -47.99 & -69.79 \\
-0.42 & 1.19 & 0.72 & -39.74 & 7.1 & -51.47 \\
0.8 & 2.8 & 1.32 & -61.71 & -69.37 & -29.09 \\
\end{bmatrix}
\mathrm{m}^2\mathrm{s}^{-1}\mathrm{Pa}^{-1}\\
\end{aligned} 
\right.
\end{equation*}
\begin{equation*}
\left.
\begin{aligned} 
& \Delta C_{ij}=
10^{-9} \times
\begin{bmatrix}
  -2.65 & -2.17 & -3.81 & 2.28 & 1.06 & 2.66
\end{bmatrix}
\mathrm{m}^2\mathrm{s}^{-1}\\
\end{aligned} 
\right.
\end{equation*}
\begin{equation*}
\left.
\begin{aligned} 
& \Delta a_{kl}=
10^{-12} \times
\begin{bmatrix}
3.18 & 4.29 & -2.48 & -9.14 & -8.02 & -2.35 \\
\end{bmatrix}   
\mathrm{eV \,Pa}^{-1}\\
\end{aligned} 
\right.
\end{equation*}
\begin{equation*}
\left.
\begin{aligned} 
& \Delta b= 
10^{-3} \times 5.85
\mathrm{eV}
\end{aligned} 
\right.
\end{equation*}

\end{document}